\documentclass[%
nofootinbib,
 amsmath,amssymb,
 aps,
prd,
eqsecnum,
notitlepage
]{revtex4-1}

\usepackage{graphicx}
\usepackage{dcolumn}
\usepackage{bm}
\usepackage{hyperref}

\newcommand{\citex}[1]{Ref.~\cite{#1}}

\newcommand{\apjl}{Astrophys. J. L.}
\newcommand{\mnras}{Mon. Not. R. Astron. Soc.}

\begin{document}


\title{Tidal deformation of a slowly rotating material body: \\ Interior metric and Love numbers}

\author{Philippe Landry}
\affiliation{%
 Department of Physics, University of Guelph, Guelph, Ontario N1G 2W1, Canada
}

\date{\today}

\begin{abstract}

The metric outside a compact body deformed by a quadrupolar tidal field is universal up to its Love numbers, constants which encode the tidal response's dependence on the body's internal structure. For a nonrotating body, the deformed external geometry is characterized by the familiar gravitational Love numbers $K_2^{\text{el}}$ and $K_2^{\text{mag}}$. For a slowly rotating body, these must be supplemented by rotational-tidal Love numbers, which measure the response to couplings between the body's spin and the external tidal field. By integrating the interior field equations, I find that the response of a barotropic perfect fluid to spin-coupled tidal perturbations is described by two rotational-tidal Love numbers, which I calculate explicitly for polytropes. Two other rotational-tidal Love numbers identified in prior work are found to have a fixed, universal value for all barotropes. Equipped with the complete interior solution, I calculate the amplitude of the time-varying internal currents induced by the gravitomagnetic part of the tidal field. For a typical neutron star in an equal-mass binary system, the size of the equatorial velocity perturbation is on the order of kilometers \mbox{per second}.
\end{abstract}

\maketitle


\section{Introduction and Summary}
\label{sec:intro}

\subsection{This work and its context}
\label{subsec:context}

The tidal response of slowly rotating bodies in general relativity has been the subject of several recent studies \cite{Poisson_TidDefRotBH, Landry_TidDefRotExt, Pani_TidDefSpin, Pani_TidRotLN, Landry_DynResp, Poisson_PNDynResp} which contribute to the broader quest of understanding the impact of tidal deformations on the gravitational wave profile of inspiraling compact binaries. Because the tidal properties of compact objects depend sensitively on their internal structure, measurements of tidal effects in the waveform could serve as a useful probe of the stellar interior. In particular, Flanagan and Hinderer have observed that measurements of the tidal phase shift in the gravitational waves emitted by neutron star binaries may be used to constrain the neutron star equation of state \cite{Flanagan_NSLNGW, Hinderer_NSLN}. This prospect has motivated the rapid development of a relativistic theory of tidal deformations \cite{Damour_TidPropNS, Binnington, Landry_SurfLN}, which accounts for both gravitoelectric and gravitomagnetic tidal fields, and which encodes a body's deformability in a set of internal-structure-dependent constants called Love numbers.

A gravitoelectric tidal field is sourced by a matter distribution far removed from the reference body; a gravitomagnetic tidal field is generated by the currents produced by this distribution. The gravitoelectric Love number $K_2^{\text{el}}$ measures the response to the quadrupole moment $\mathcal{E}_{ab}$ of the gravitoelectric tidal field; the gravitomagnetic Love number $K_2^{\text{mag}}$ does likewise for the quadrupole moment $\mathcal{B}_{ab}$ of the gravitomagnetic tidal field. Together, they provide a complete description of the tidal deformation of a nonrotating body at leading order in the tidal interaction. The gravitational Love numbers $K_2^{\text{el}}$ and $K_2^{\text{mag}}$ have been computed for model neutron stars with realistic equations of state and have been incorporated in analytically and numerically constructed neutron star binary waveforms \cite{Hinderer_TidDefNS, Baiotti_TidEffInsp, Baiotti_InspBinNS, Vines_PNTidEff, Pannarale, Lackey_NonSpinBH, Damour_TidPolarNS, Read_MatterEffGW, Vines_PNQuadTid, Maselli_NSEoSGW, Lackey_AlignedSpinBH, Favata_ParamErr, Yagi_HardToMeasure}. They have also been implicated in the I-Love-Q universality relations \cite{Yagi_ILoveQ, Yagi_ILoveQApp, Delsate_ILove, Doneva, Maselli_EoSIndepRelNS, Yagi_MultiLove, Haskell, Chakrabarti_IQRapidRotNS}, and have been shown to vanish for black holes \cite{Binnington}.

This paper is concerned with the case of a rotating body, which is complicated by couplings between the body's angular momentum and the tidal field; at first order in the (dimensionless) spin $\chi^a$, four additional Love numbers are required to fully describe the deformation. These new \emph{rotational-tidal Love numbers} were introduced by Landry and Poisson in \citex{Landry_TidDefRotExt} (Paper I): $\mathfrak{E}^{\text{q}}$ and $\mathfrak{F}^{\text{o}}$ measure the respective quadrupolar ($\mathrm{q}$) and octupolar ($\mathrm{o}$) deformations induced by the coupling between $\chi^a$ and $\mathcal{E}_{ab}$, while $\mathfrak{B}^{\text{q}}$ and $\mathfrak{K}^{\text{o}}$ measure the respective quadrupolar and octupolar deformations that result from the coupling between $\chi^a$ and $\mathcal{B}_{ab}$. A different, but related, set of rotational-tidal Love numbers was introduced by Pani \textit{et al.}~in \citex{Pani_TidDefSpin}. The technical differences between their formalism and the one employed here are discussed in Sec.~\ref{subsec:Pani}.

The metric outside a tidally deformed, slowly rotating black hole was constructed by Poisson in \citex{Poisson_TidDefRotBH}. In Paper I, the black hole was replaced with a material body, and gravitational and rotational-tidal Love numbers appeared in the exterior metric. The metric of Paper I reduces to that of \citex{Poisson_TidDefRotBH} when the Love numbers are set to zero, indicating that black holes have vanishing rotational-tidal Love numbers. For a material body, the values of the Love numbers depend on the details of internal structure; to compute them it is necessary to solve the field and fluid equations in the body's interior. The calculation of the rotational-tidal Love numbers for material bodies is the primary objective of this paper. A portion of the task was already accomplished in \citex{Landry_DynResp} (Paper II), which presented---but did not solve---the interior field equations governing perturbations generated by a spin-coupled gravitomagnetic tidal field. In this paper, the field equations which were omitted in Paper II---those which govern spin-coupled gravitoelectric tidal perturbations---are examined. Then, the full set of interior field equations is integrated to determine the rotational-tidal Love numbers explicitly for polytropes. Before going on to describe these calculations in detail, I outline the setup of the problem, and summarize the key findings of the paper.

I consider a material body of mass $M$, radius $R$ and rotational angular velocity vector $\Omega^a$ immersed in a tidal environment characterized by the gravitoelectric tidal (quadrupole) moment $\mathcal{E}_{ab}$ and the gravitomagnetic tidal (quadrupole) moment $\mathcal{B}_{ab}$. The details of the construction of the tidal environment are provided in Paper I; I recapitulate them briefly here. I work in the regime of stationary tides, assuming that the body's response occurs rapidly compared to the time scale for variation of the tidal environment. This regime is characteristic of the inspiral phase of a binary system, since the wide separation $b \gg R$ of the orbiting bodies means the orbital period is much longer than the rotational period of the reference body. In this setting, the body's response takes place over the short hydrodynamical time scale $(R^3/M)^{1/2}$, while the phase of the tidal moments varies on the long orbital time scale $(b^3/M)^{1/2}$ and their amplitude varies on the even longer radiation-reaction time scale. For my purposes, I neglect the changes in the tidal environment and take $\mathcal{E}_{ab}$ and $\mathcal{B}_{ab}$ to be time independent, but otherwise generic.

To determine the solution in the body's interior, which completes the exterior solution given in Eq.~(4.4) of Paper I, one must solve the Einstein field equations for the metric together with the relativistic Euler equation for the fluid variables. This, in turn, requires a specification of the fluid's state. Following Paper II, I assume that the body is made up of a perfect fluid which satisfies a barotropic equation of state, and that the Lagrangian perturbation $\Delta \omega_{\alpha\beta}$ of the vorticity tensor vanishes throughout the evolution of the fluid. This is the natural state that arises if one assumes that the barotropic body was isolated in the remote past, before the tidal field was switched on adiabatically. I shall refer to this state as ``irrotational'' since, at zeroth order in spin, it permits the establishment of vorticity-free internal fluid motions via gravitomagnetic induction (cf.~\citex{Landry_Irrot}). The irrotational state stands in contrast to the ``static'' state typically employed in work on tidal deformations (e.g.~Refs.~\cite{Damour_TidPropNS, Binnington, Pani_TidDefSpin, Pani_TidRotLN}), in which the fluid is held in a strict hydrostatic equilibrium that prevents any motion. In Paper II, Landry and Poisson showed that a subset of the internal metric and fluid variables associated with the gravitomagnetic response of a fluid body in the irrotational state must be time dependent, even when $\mathcal{B}_{ab}$ is stationary. In contrast, here I find that the gravitoelectric response bears no trace of internal dynamics.

\begin{figure}
\includegraphics[width=0.75\textwidth]{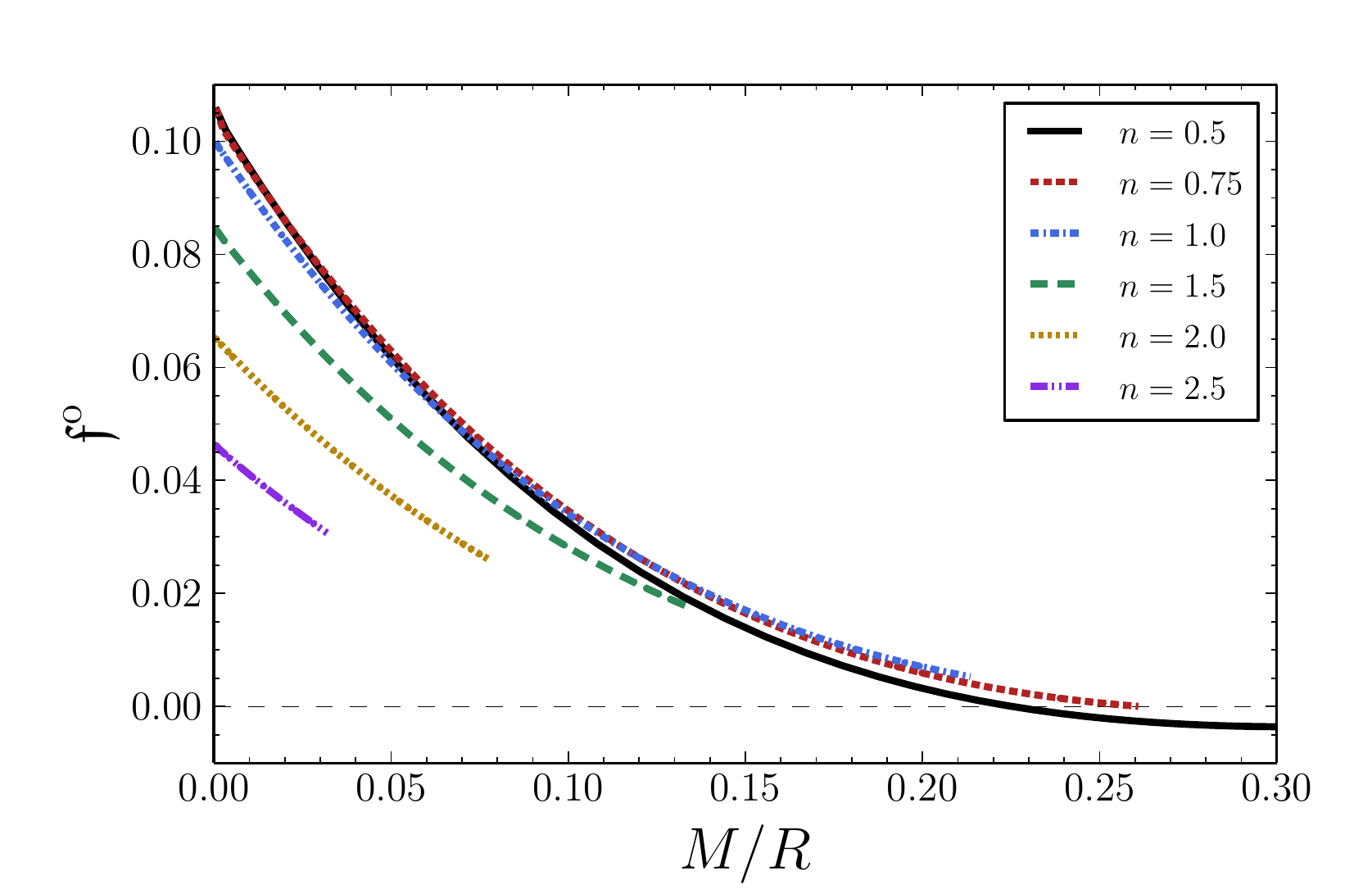}
\caption{\label{fig:fo} Scale-free gravitoelectric rotational-tidal Love number $\mathfrak{f}^{\text{o}} = -(2M/R)^5 \mathfrak{F}^{\text{o}}$ as a function of compactness $M/R$ for polytropes of index $n$. The Love numbers are computed up to the maximum compactness supported by the given \mbox{equation of state}.}
\end{figure}

\begin{figure}
\includegraphics[width=0.75\textwidth]{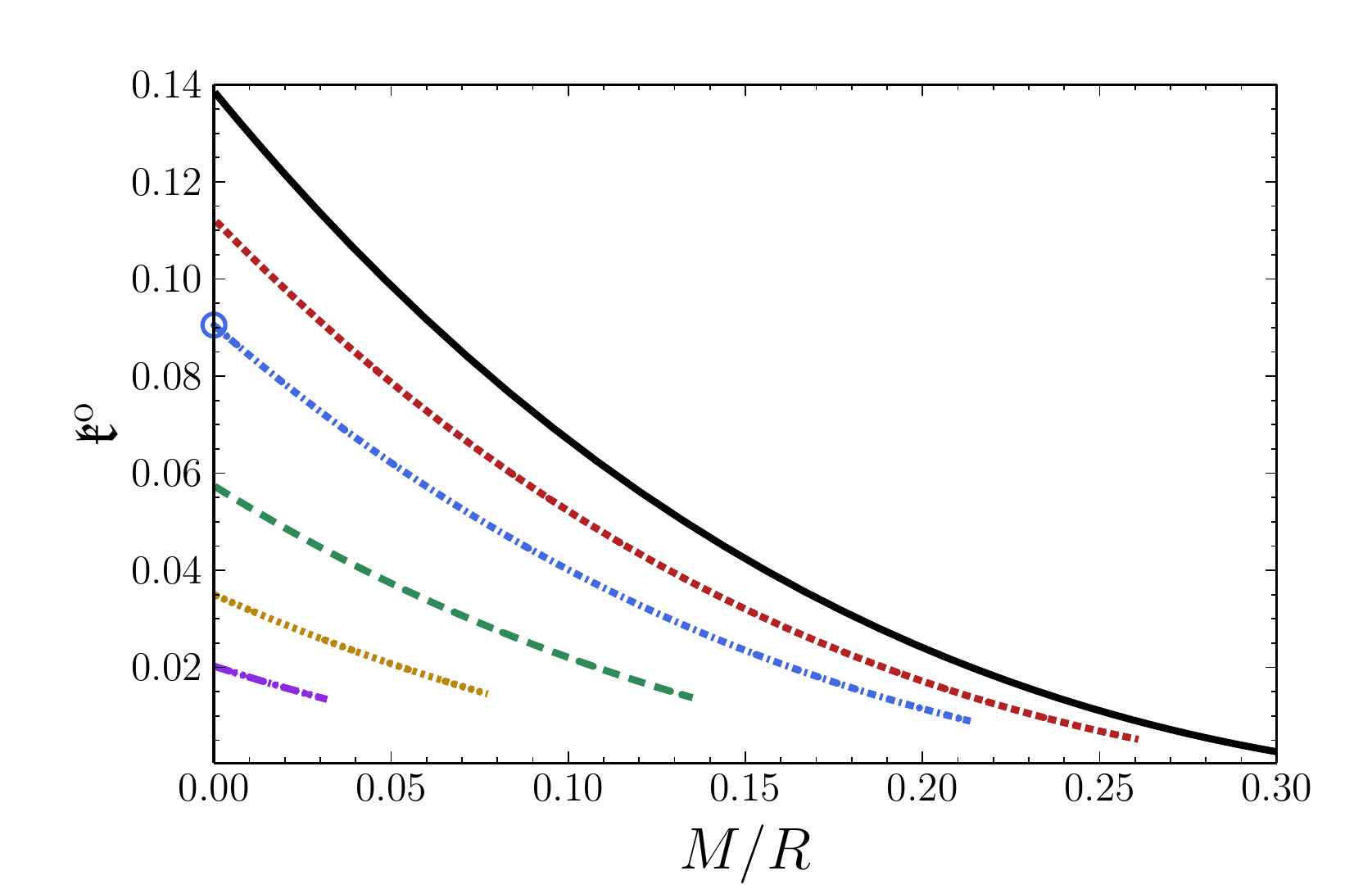}
\caption{\label{fig:ko} Scale-free gravitomagnetic rotational-tidal Love number $\mathfrak{k}^{\text{o}} = - (2M/R)^5 \mathfrak{K}^{\text{o}}$ as a function of compactness $M/R$ for polytropes of index $n$. The Love numbers are computed up to the maximum compactness supported by the given equation of state. The circled data point at $M/R = 0$ represents the post-Newtonian result of \citex{Poisson_PNDynResp} for an $n=1$ polytrope.}
\end{figure}

Combining the solutions to the field equations for the gravitomagnetic sector treated in Paper II and the gravitoelectric sector treated here, I match the complete interior metric to the exterior metric of Paper I at the body's surface, and determine the Love numbers appearing therein. I find that the response of a slowly rotating material body to spin-coupled tidal fields is measured by just two rotational-tidal Love numbers, $\mathfrak{F}^{\text{o}}$ and $\mathfrak{K}^{\text{o}}$. They scale with the body's compactness like $\mathfrak{F}^{\text{o}} \sim \mathfrak{f}^{\text{o}} (R/M)^5$ and $\mathfrak{K}^{\text{o}} \sim \mathfrak{k}^{\text{o}} (R/M)^{5} $; the scaling of $\mathfrak{F}^{\text{o}}$ was foreseen in Paper I, and the scaling of $\mathfrak{K}^{\text{o}}$ was predicted by \citex{Poisson_PNDynResp}. The dimensionless rotational-tidal Love numbers $\mathfrak{f}^{\text{o}}$ and $\mathfrak{k}^{\text{o}}$ are plotted as a function of compactness in Fig.~\ref{fig:fo} and Fig.~\ref{fig:ko}, respectively, for various polytropic equations of state. The rotational-tidal Love number $\mathfrak{k}^{\text{o}}$ decreases monotonically in magnitude with increasing compactness, like the gravitational Love numbers for polytropes (cf.~Refs.~\cite{Binnington, Landry_Irrot}). In the limit of zero compactness, the value of $\mathfrak{k}^{\text{o}}$ for a polytrope of index $n=1$ matches the result of the post-Newtonian calculation of \citex{Poisson_PNDynResp}. The rotational-tidal Love number $\mathfrak{f}^{\text{o}}$ also decreases in magnitude as $M/R$ increases; however, it changes sign from positive to negative along a sequence of increasing compactness for sufficiently stiff equations of state. The sign of the Love number reflects whether the tide has a stretching or compressing effect on the body. The existence of a compressive component in relativistic tides was first pointed out by Shapiro \cite{Shapiro}, and has been discussed in \citex{Favata_CrushedNS}.

The other two rotational-tidal Love numbers, $\mathfrak{E}^{\text{q}}$ and $\mathfrak{B}^{\text{q}}$, are found to have a fixed, universal value for all material bodies. $\mathfrak{E}^{\text{q}}$ appears in the external metric variables $\lbrace \hat{e}^\text{q}_{tt}, \hat{e}^\text{q}_{rr}, \hat{e}^\text{q} \rbrace$ listed in Table IV of Paper I, while $\mathfrak{B}^{\text{q}}$ appears in $\hat{b}^{\text{q}}_t$; the assignment $\mathfrak{E}^{\text{q}} = 1/120 = \mathfrak{B}^{\text{q}}$ dictated by the field equations causes each of these metric variables to vanish. This means that the metric outside a tidally deformed, slowly rotating material body takes on the simplified form

\begin{subequations} \label{extmetric}
\begin{eqnarray}
g_{tt} &=& -(1-2M/r) + e^{\text{q}}_{tt} \mathcal{E}^{\text{q}} + k^{\text{d}}_{tt} \mathcal{K}^{\text{d}} + k^{\text{o}}_{tt} \mathcal{K}^{\text{o}} , \\
g_{tr} &=& \hat{e}^{\text{q}}_{tr} \hat{\mathcal{E}}^{\text{q}} , \\
g_{rr} &=& (1-2M/r)^{-1} + e^{\text{q}}_{rr} \mathcal{E}^{\text{q}} + k^{\text{d}}_{rr} \mathcal{K}^{\text{d}} + k^{\text{o}}_{rr} \mathcal{K}^{\text{o}} , \\
g_{tA} &=& \frac{2 M^2}{r}\chi^{\text{d}}_A + b^{\text{q}}_{t} \mathcal{B}^{\text{q}}_A + f^{\text{d}}_{t} \mathcal{F}^{\text{d}}_A + f^{\text{o}}_{t} \mathcal{F}^{\text{o}}_A , \\
g_{rA} &=& \hat{b}^{\text{q}}_{r} \hat{\mathcal{B}}^{\text{q}}_A , \\
g_{AB} &=& r^2 \Omega_{AB} + e^{\text{q}} \Omega_{AB} \mathcal{E}^{\text{q}} + k^{\text{o}} \Omega_{AB} \mathcal{K}^{\text{o}}
\end{eqnarray}
\end{subequations}
in Boyer-Lindquist ($t,r,\theta,\phi$) coordinates, where uppercase latin indices $A,B,C, ...$ range over the angular variables ($\theta,\phi$) and $\Omega_{AB} = \text{diag}(1,\sin^2{\theta})$ is the two-sphere metric. The tidal potentials $\lbrace \mathcal{E}^{\text{q}}, \mathcal{K}^{\text{d}}, ..., \mathcal{K}^{\text{o}} \rbrace$ are defined in Sec.~II of Paper I, and the radial functions $\lbrace e^{\text{q}}_{tt}, k^{\text{d}}_{tt}, ..., k^{\text{o}} \rbrace$ are listed in Table IV of Paper I (subject to the change of notation described in footnotes 1 and 3, and footnotes 4 and 6 of Paper II). The metric in Eq.~\eqref{extmetric} is universal up to the gravitational Love numbers $K_2^{\text{el}}$ and $K_2^{\text{mag}}$ which appear in $\lbrace e^{\text{q}}_{tt}, e^{\text{q}}_{rr}, e^{\text{q}} \rbrace$ and $b^{\text{q}}_t$, respectively, and the rotational-tidal Love numbers $\mathfrak{F}^{\text{o}}$ and $\mathfrak{K}^{\text{o}}$ which appear in $f^{\text{o}}_t$ and $\{ k^{\text{o}}_{tt}, k^{\text{o}}_{rr}, k^{\text{o}} \}$, respectively.

The plan of the paper is as follows: I present the background metric of a static, spherically symmetric body and set up the rotational, tidal and bilinear perturbations in Sec.~\ref{sec:spacetime}. The gravitomagnetic field $\mathcal{B}_{ab}$ is left out of the construction altogether, since Paper II already treated spin-coupled gravitomagnetic perturbations. In Sec.~\ref{sec:fluid}, I introduce the fluid variables and work out the consequences of the vorticity conservation condition $\Delta \omega_{\alpha\beta} = 0$. In Sec.~\ref{sec:efe}, a complete set of equations for all internal metric and fluid variables associated with the gravitoelectric field $\mathcal{E}_{ab}$ is obtained from the Einstein field equations. The field equations are integrated, and the interior solution is matched to the exterior solution of Paper I to determine the rotational-tidal Love numbers $\mathfrak{E}^{\text{q}}$ and $\mathfrak{F}^{\text{o}}$. Finally, in Sec.~\ref{sec:gravitomag}, I revisit the field equations from Sec.~IV of Paper II governing the gravitomagnetic sector of the tidal response to determine the rotational-tidal Love numbers $\mathfrak{B}^{\text{q}}$ and $\mathfrak{K}^{\text{o}}$. Having established the complete interior solution, I calculate the amplitude of the gravitomagnetically induced dynamical internal currents identified in Paper II. For a neutron star in a binary system of relevance to LIGO, the size of the tangential velocity perturbation is over a kilometer per second at the equator.

\subsection{Pani, Gualtieri and Ferrari}
\label{subsec:Pani}

The objective of this paper was pursued in parallel by Pani, Gualtieri and Ferrari in \citex{Pani_TidRotLN}. I shall briefly clarify the differences between their work and mine. Just as this paper depends on the identification of the rotational-tidal Love numbers made in Paper I, \citex{Pani_TidRotLN} relies on the formalism of a prior paper by Pani \textit{et al.}, \citex{Pani_TidDefSpin}, in which the external geometry of a tidally deformed, rotating star is calculated to second order in its spin. Because the formalism of \citex{Pani_TidDefSpin} is adapted to axisymmetric settings, \citex{Pani_TidRotLN} lacks the analogues of my Love numbers $\mathfrak{E}^{\text{q}}$ and $\mathfrak{B}^{\text{q}}$; however, since \citex{Pani_TidDefSpin} treats the response to the tidal field's higher multipole moments, \citex{Pani_TidRotLN} has additional Love numbers relative to this paper. Nevertheless, both \citex{Pani_TidRotLN} and this work compute the octupolar rotational-tidal Love numbers which measure the response to spin-coupled quadrupolar tidal fields. By comparing the external metric of \citex{Pani_TidRotLN} to the one presented in Paper I, it is possible to establish the mapping between the definitions of the Love numbers in each formalism. I find that the Love numbers $\delta \tilde{\lambda}_M^{(32)}$ and $\delta \tilde{\lambda}_E^{(32)}$ of \citex{Pani_TidRotLN} can be expressed as a linear combination of my gravitational and rotational-tidal Love numbers; specifically,

\begin{equation} \label{map}
\delta \tilde{\lambda}_M^{(32)} = -\frac{32}{\sqrt{5\pi}} \left( 5 K_2^{\text{el}} - 3 \mathfrak{F}^{\text{o}} \right) , \qquad \delta \tilde{\lambda}_E^{(32)} \propto -144\sqrt{\frac{7}{5}} \mathfrak{K}^{\text{o}} .
\end{equation}
The mapping has been established modulo a prefactor in the latter case since $\delta \tilde{\lambda}_E^{(32)}$ is defined only up to an overall scale in \citex{Pani_TidRotLN}.

Pani, Gualtieri and Ferrari compute rotational-tidal Love numbers for polytropes of index $n=1$ and tabulated neutron star equations of state. Since I restrict my attention to polytropic equations of state in this paper, our results coincide only in the former case. Even so, the fact that we make different physical assumptions about the material body means that we cannot expect our results to be in quantitative agreement. First and foremost, Pani, Gualtieri and Ferrari place the material body in the static fluid state, which artificially prevents internal currents from developing through gravitomagnetic induction. As shown in Paper II, the static state is compatible with the Einstein field equations only in axisymmetry; the irrotational state employed in this paper is valid in generic settings. The numerical value of the Love numbers depends on the choice of fluid state (cf.~\citex{Landry_Irrot}), and our results will therefore differ in this respect.

Second, the polytropic model adopted in \citex{Pani_TidRotLN} is different from the one employed here. Pani, Gualtieri and Ferrari use the equation of state

\begin{equation} \label{epoly}
p = K \mu^{1+1/n} ,
\end{equation}
which describes an ``energy polytrope'' whose pressure $p$ is related to the total energy density $\mu$ (the sum of mass density $\rho$ and internal energy density $\epsilon$). $K$ is a constant of proportionality which sets the polytrope's compactness $M/R$ for a given choice of $n$ and central density $\mu_c := \mu(r=0)$. In contrast, my chosen polytropic equation of state is

\begin{equation} \label{mpoly}
p = K \rho^{1+1/n} ,
\end{equation}
which describes a ``mass polytrope'' whose pressure is related to the mass density $\rho$, and whose total energy density is $\mu = \rho + n p$. Tooper \cite{Tooper_AdFluidSph} has shown that Eq.~\eqref{mpoly}, unlike Eq.~\eqref{epoly}, always produces a sound speed less than unity for $n \geq 1$. The difference between the two equations of state becomes more pronounced for larger values of the compactness.

Because of these significant differences in formulation, I will not attempt a precise comparison of numerical results with Pani, Gualtieri and Ferrari in this paper. However, I do note that when subjected to the mapping prescribed in Eq.~\eqref{map}, my results are similar in magnitude to those of Pani, Gualtieri and Ferrari, and display the same qualitative behavior as a function of the compactness as can be seen in Figs.~5 and 9 of \citex{Pani_TidRotLN}.

\section{Spacetime of a Tidally Deformed, Slowly Rotating Body}
\label{sec:spacetime}

The metric inside a tidally deformed, slowly rotating perfect fluid body is built up from successive perturbations of the background metric of an isolated, nonrotating body in hydrostatic equilibrium,

\begin{equation}
ds^2 = - e^{2\psi} dt^2 + f^{-1} dr^2 + r^2 d\Omega^2 ,
\end{equation}
where $d\Omega^2 := d\theta^2 + \sin^2{\theta}\,d\phi^2$. The functions $\psi(r)$ and $f = 1-2m(r)/r$ are determined by the Einstein field equations

\begin{equation}
\frac{dm}{dr} = 4 \pi r^2 \mu , \qquad \frac{d\psi}{dr} = \frac{m + 4\pi r^3 p}{r^2 f} . \label{mass}
\end{equation}
The fluid's pressure $p$ satisfies the condition of hydrostatic equilibrium

\begin{equation}
\frac{dp}{dr} = - \frac{(\mu + p)(m+ 4\pi r^3 p)}{r^2 f} , \label{hydro}
\end{equation}
and its total energy density $\mu$ is the sum of mass density $\rho$ and internal energy density $\epsilon$. The fluid's equation of state is taken to be barotropic, such that $p=p(\rho)$ and $\epsilon = \epsilon(\rho)$. The interior metric matches on to the exterior Schwarzschild solution at the body's surface $r=R$, where $p=0$.

The body's slow, rigid rotation is added as a linear, dipole perturbation of the background metric;

\begin{equation}
p_{t\phi}^{\text{rotation}} = - \Omega(1-\omega) r^2 \sin^2{\theta}
\label{rotpert}
\end{equation}
is its only nonzero component. The body's angular velocity is $\Omega$, and the function $\omega(r)$ satisfies \cite{FriedmanStergioulas}

\begin{equation}
r^2 f \frac{d^2 \omega}{ dr^2} + \left[ 4f-4\pi r^2 (\mu+p) \right] r \frac{d \omega}{dr} - 16 \pi r^2 (\mu + p) \omega = 0 .
\end{equation}
At $r=R$, $\omega$ matches the exterior solution $\omega_{\text{ext}} = 1- 2I/r^3$, where $I := \chi M^2/\Omega$ is the body's moment of inertia.

The gravitoelectric tidal field which deforms the body is characterized by a time-independent quadrupole moment $\mathcal{E}_{ab}$. This Cartesian tensor is symmetric tracefree (STF) and has five independent components which are conveniently packaged in spherical-harmonic coefficients $\mathcal{E}^{\text{q}}_{\mathsf{m}}$. It is likewise convenient to decompose the tidal perturbation in spherical harmonics. Since $\mathcal{E}_{ab}$ generates polar (even-parity) perturbations, the decomposition involves the scalar harmonics $Y^{\ell \mathsf{m}}$ and the even-parity vector harmonics

\begin{equation}
Y^{\ell {\mathsf{m}}}_A := D_A Y^{\ell {\mathsf{m}}} ,
\end{equation}
where $D_A$ is the covariant derivative operator on the unit two-sphere. I adopt the normalization given in Table II of Paper I for the spherical harmonics; the corresponding coefficients $\mathcal{E}^{\text{q}}_{\mathsf{m}}$ are listed in Table I of Paper I.

Expressed in the Regge-Wheeler gauge, the tidal perturbation's nonvanishing components are\footnote{My notation for the radial functions differs from the one adopted in Sec.~IV of Paper I; specifically, $e^{\text{q}}_{tt}[\text{here}] = -r^2 e^{\text{q}}_{tt}[\text{Paper I}]$, $e^{\text{q}}_{rr}[\text{here}] = -r^2 e^{\text{q}}_{rr}[\text{Paper I}]$ and $e^{\text{q}}[\text{here}] = -r^4 e^{\text{q}}[\text{Paper I}]$.}

\begin{subequations}
\label{tidpert}
\begin{align}
p_{tt}^{\text{tidal}} &= e^{\text{q}}_{tt}(r) \mathcal{E}^{\text{q}} , \\
p_{rr}^{\text{tidal}} &= e^{\text{q}}_{rr}(r) \mathcal{E}^{\text{q}} , \\
p_{AB}^{\text{tidal}} &= e^{\text{q}}(r) \Omega_{AB} \mathcal{E}^{\text{q}} ,
\end{align}
\end{subequations}
with

\begin{equation}
\mathcal{E}^{\text{q}} := \sum_{\mathsf{m}} \mathcal{E}^{\text{q}}_{\mathsf{m}} Y^{2 \mathsf{m}} .
\end{equation}
The radial function $e^{\text{q}}_{tt}$ satisfies the field equation

\begin{equation}
r^2 f \frac{d^2 e^{\text{q}}_{tt}}{dr^2} - 2 \left[ \frac{3m}{r} - 1 + 2 \pi  r^2 (\mu + 3p) \right] r \frac{d e^{\text{q}}_{tt}}{dr} - 2 \left[3-2 \pi  r^2 (\mu + p)\left(3+\frac{d\mu}{dp}\right) \right] e^{\text{q}}_{tt} = 0 ; \label{eqtt}
\end{equation}
the other radial functions are related by

\begin{equation}
e^{\text{q}}_{rr} = f^{-1} e^{-2\psi} e^{\text{q}}_{tt} \label{eqrr}
\end{equation}
and

\begin{equation}
e^{\text{q}} = r^2 e^{-2 \psi} \left\lbrace \frac{1}{2} \left(\frac{m}{r}+4 \pi r^2 p\right) r \frac{d e^{\text{q}}_{tt}}{dr} + \left[1 + \frac{m}{r}+4 \pi r^2 p - 2 \pi  r^2 (\mu + 3 p) \right] e^{\text{q}}_{tt} \right\rbrace . \label{eq}
\end{equation}
At $r=R$, each of the variables $\lbrace e^{\text{q}}_{tt}, e^{\text{q}}_{rr}, e^{\text{q}} \rbrace$ matches its external expression from Table I of Paper I.

Second-order perturbations are sourced by the (dipole) rotational perturbation of Eq.~\eqref{rotpert} and the (quadrupole) tidal perturbation of Eq.~\eqref{tidpert}. I continue to work to first order in each of $\Omega$ and $\mathcal{E}_{ab}$, but now introduce terms of order $\Omega \mathcal{E}_{ab}$ in the perturbed metric. The composition of the $\ell = 1$ and $\ell = 2$ spherical harmonics is reflected in the bilinear moments\footnote{I use the same notation as in Paper I for the bilinear moments, but construct them with $\Omega^a$ instead of $\chi^a$. Consequently, the moments differ by a factor of $\chi/\Omega = I/M^2$. \label{chiomega}}

\begin{eqnarray}
\mathcal{F}_a := \mathcal{E}_{ab} \Omega^b , \qquad \hat{\mathcal{E}}_{ab} := 2 \Omega^c \epsilon_{cd(a} \mathcal{E}^{\;\,d}_{b)} , \qquad \mathcal{F}_{abc} := \mathcal{E}_{\langle ab} \Omega_{c \rangle} ,
\end{eqnarray}
where $\Omega_a := [0,0,\Omega]$ is the angular velocity vector, $\epsilon_{abc}$ is the antisymmetric permutation symbol, and angular brackets indicate the STF operation (symmetrize all indices and remove all traces). Like $\mathcal{E}_{ab}$, the independent components of the STF tensors $\mathcal{F}_a, \hat{\mathcal{E}}_{ab}$ and $\mathcal{F}_{abc}$ can be packaged in spherical harmonic coefficients $\mathcal{F}^{\text{d}}_{\mathsf{m}}, \hat{\mathcal{E}}^{\text{q}}_{\mathsf{m}}$ and $\mathcal{F}^{\text{o}}_{\mathsf{m}}$, respectively; the precise packaging is presented in Table I of Paper I (with the change of notation described in footnote \ref{chiomega}). The perturbations created by $\mathcal{F}_a$ and $\mathcal{F}_{abc}$ are axial (odd-parity) in nature, while those created by $\hat{\mathcal{E}}_{ab}$ are polar (even-parity).

The second-order perturbation is also expressed in the Regge-Wheeler gauge, with the extension of Campolattaro and Thorne \cite{Campolattaro} for the $\ell = 1$ terms generated by $\mathcal{F}_a$. Its components are\footnote{My notation here also differs from Sec.~IV of Paper I: $\hat{e}^{\text{q}}_{tt}[\text{here}] = r^2 \hat{e}^{\text{q}}_{tt}[\text{Paper I}]$, $\hat{e}^{\text{q}}_{tr}[\text{here}] = r^2 \hat{e}^{\text{q}}_{tr}[\text{Paper I}]$, $\hat{e}^{\text{q}}_{rr}[\text{here}] = r^2 \hat{e}^{\text{q}}_{rr}[\text{Paper I}]$, $f^{\text{d}}_t[\text{here}] = -r^3 f^{\text{d}}_t[\text{Paper I}]$, $f^{\text{o}}_t[\text{here}] = r^3 f^{\text{o}}_t[\text{Paper I}]$, and $\hat{e}^{\text{q}}[\text{here}] = r^4 \hat{e}^{\text{q}}[\text{Paper I}]$.}

\begin{subequations}
\label{ansatz}
\begin{eqnarray}
p_{tt}^{\text{bilinear}} &=& \hat{e}^{\text{q}}_{tt}(t,r) \hat{\mathcal{E}}^{\text{q}} , \\
p_{tr}^{\text{bilinear}} &=& \hat{e}^{\text{q}}_{tr}(t,r) \hat{\mathcal{E}}^{\text{q}} , \\
p_{rr}^{\text{bilinear}} &=& \hat{e}^{\text{q}}_{rr}(t,r) \hat{\mathcal{E}}^{\text{q}} , \\
p_{tA}^{\text{bilinear}} &=& f^{\text{d}}_t(t,r) \mathcal{F}^{\text{d}}_A + f^{\text{o}}_t(t,r) \mathcal{F}^{\text{o}}_A , \\
p_{rA}^{\text{bilinear}} &=& f^{\text{o}}_r(t,r) \mathcal{F}^{\text{o}}_A , \\
p_{AB}^{\text{bilinear}} &=& \hat{e}^{\text{q}}(t,r) \Omega_{AB} \hat{\mathcal{E}}^{\text{q}} ,
\end{eqnarray}
\end{subequations}
where

\begin{eqnarray}
\mathcal{F}^{\text{d}}_A := \sum_{\mathsf{m}} \mathcal{F}^{\text{d}}_{\mathsf{m}} X^{1\mathsf{m}}_A , \qquad \mathcal{F}^{\text{o}}_A := \frac{1}{3} \sum_{\mathsf{m}} \mathcal{F}^{\text{o}}_{\mathsf{m}} X^{3\mathsf{m}}_A , \qquad \hat{\mathcal{E}}^{\text{q}} := \sum_{\mathsf{m}} \hat{\mathcal{E}}^{\text{q}}_{\mathsf{m}} Y^{2\mathsf{m}} = - \Omega \partial_{\phi} \mathcal{E}^{\text{q}} ,
\end{eqnarray}
and

\begin{equation}
X_A^{\ell \mathsf{m}} := - \epsilon_A^{\;\; B} D_B Y^{\ell \mathsf{m}}
\end{equation}
are the odd-parity vector harmonics, with $\epsilon_{AB}$ the Levi-Civita tensor on the unit two-sphere ($\epsilon_{\theta \phi} = \sin{\theta}$). I allow the functions $\lbrace \hat{e}^{\text{q}}_{tt}, \hat{e}^{\text{q}}_{tr}, ..., \hat{e}^{\text{q}} \rbrace$ to depend on both $t$ and $r$, and note that $\hat{\mathcal{E}}^{\text{q}}$ vanishes when the tidal environment is axisymmetric.

\section{Perturbed Fluid}
\label{sec:fluid}

The perfect fluid which makes up the body is disturbed by the perturbations created by $\mathcal{E}_{ab}, \mathcal{F}_a, \hat{\mathcal{E}}_{ab}$, and $\mathcal{F}_{abc}$. I give this disturbance a Lagrangian formulation, and suppose that the perturbed fluid's one-parameter equation of state is the same as the unperturbed fluid's.

The Lagrangian displacement vector $\xi_{\alpha}$ describes how the fluid elements are translated by the perturbation. It is decomposed as

\begin{subequations}
\begin{eqnarray}
\xi_r &=& \xi^{\text{q}}_r(t,r) \mathcal{E}^{\text{q}} + \hat{\xi}^{\text{q}}_r(t,r) \hat{\mathcal{E}}^{\text{q}} , \label{xir} \\
\xi_A &=& \xi^{\text{q}}(t,r) \mathcal{E}^{\text{q}}_A + \xi^{\text{d}}(t,r) \mathcal{F}^{\text{d}}_A + \hat{\xi}^{\text{q}}(t,r) \mathcal{\hat{E}}^{\text{q}}_A + \xi^{\text{o}}(t,r) \mathcal{F}^{\text{o}}_A ,
\end{eqnarray}
\end{subequations}
where

\begin{equation}
\mathcal{E}^{\text{q}}_A := \frac{1}{2} \sum_{\mathsf{m}} \mathcal{E}^{\text{q}}_{\mathsf{m}} Y^{2\mathsf{m}}_A , \qquad \hat{\mathcal{E}}^{\text{q}}_A := \frac{1}{2} \sum_{\mathsf{m}} \hat{\mathcal{E}}^{\text{q}}_{\mathsf{m}} Y^{2\mathsf{m}}_A  .
\end{equation}
The time component of $\xi_{\alpha}$ is irrelevant for my purposes. The Eulerian perturbation of the velocity vector is expressed as

\begin{subequations} \label{deltau}
\begin{eqnarray}
\delta u_r &=& v^{\text{q}}_r(r) \mathcal{E}^{\text{q}} + \hat{v}^{\text{q}}_r(t,r) \hat{\mathcal{E}}^{\text{q}} , \\
\delta u_A &=& v^{\text{q}}(r) \mathcal{E}^{\text{q}}_A + v^{\text{d}}(t,r) \mathcal{F}^{\text{d}}_A + \hat{v}^{\text{q}}(t,r) \hat{\mathcal{E}}^{\text{q}}_A + v^{\text{o}}(t,r) \mathcal{F}^{\text{o}}_A ; \label{deltauA}
\end{eqnarray}
\end{subequations}
$\delta u_t$ can be related to the other components by properly normalizing the perturbed velocity vector. The Lagrangian change in $u_{\alpha}$ is expressible in terms of the Eulerian change; indeed, for a given fluid variable $Q$, the Lagrangian perturbation $\Delta Q$ and the Eulerian perturbation $\delta Q$ are related by $\Delta Q = \delta Q + \mathcal{L}_{\xi} Q$, where $\mathcal{L}_{\xi}$ is a Lie derivative with respect to $\xi^{\alpha}$ \cite{FriedmanStergioulas}.

The Eulerian perturbation of the pressure is decomposed as

\begin{equation} \label{deltap}
\delta p = p^{\text{q}}(r) \mathcal{E}^{\text{q}} + \hat{p}^{\text{q}}(t,r) \mathcal{\hat{E}}^{\text{q}} ,
\end{equation}
and the perturbations in energy density $\mu$ and specific enthalpy $h$ are given by $\delta \mu = (d\mu/dp) \delta p$ and $\delta h = h (\mu+p)^{-1} \delta p$ on account of the barotropic assumption. As a consequence of the conservation equation $\nabla_{\alpha} T^{\alpha}_{\;  \; \; \beta} = 0$ for a barotropic fluid, the vorticity tensor $\omega_{\alpha\beta} := \nabla_{\alpha}(h u_{\beta}) - \nabla_{\beta}(h u_{\alpha})$ is conserved along the fluid worldlines: $\mathcal{L}_u \omega_{\alpha\beta} = 0$ (see \citex{FriedmanStergioulas} for a derivation of this result). The perturbed version of this statement, $\mathcal{L}_u \Delta \omega_{\alpha\beta} = 0$, must hold for the perturbed fluid, which means that $\Delta \omega_{\alpha\beta}$ is conserved along the fluid worldlines. As explained in Paper II, if one supposes---despite the assumption of stationarity---that the tidal perturbation was switched on adiabatically in the distant past (so that $\Delta \omega_{\alpha\beta} = 0$ in the unperturbed initial state), the conservation statement ensures that

\begin{equation}
\Delta \omega_{\alpha\beta} = 0
\label{vortcons}
\end{equation}
at all times.

The vorticity conservation condition constrains several of the fluid variables introduced in Eqs.~\eqref{deltau} and \eqref{deltap}. The $tA$ components of Eq.~\eqref{vortcons} produce the assignments

\begin{subequations}
\begin{align}
p^{\text{q}} &= \frac{1}{2} e^{-2\psi} (\mu+p) e^{\text{q}}_{tt} , \label{pq} \\
\hat{p}^{\text{q}} &= \frac{1}{2} e^{-\psi} (\mu+p) (e^{-\psi} \hat{e}^{\text{q}}_{tt} + v^{\text{q}}) . \label{phatq}
\end{align}
\end{subequations}
The $rA$ components then yield

\begin{equation}
v^{\text{q}}_r = \frac{1}{2} \left( \frac{d v^{\text{q}}}{dr} - \frac{m + 4\pi r^3 p}{r^2 f} v^{\text{q}} \right) . \label{vqr}
\end{equation}
The $rA$ components also relate $\hat{v}^{\text{q}}_r$ to $\hat{v}^{\text{q}}$, $\xi^{\text{q}}$ and $\xi^{\text{q}}_r$; however Eq.~\eqref{vortcons} does not provide enough information to specify all four of these fluid variables. Consequently, I leave the determination of $\hat{v}^{\text{q}}_r$ and $\hat{v}^{\text{q}}$ up to Sec.~\ref{sec:efe}'s analysis of the Einstein field equations.\footnote{In fact, I will show that $\hat{v}^{\text{q}}_r$ and $\hat{v}^{\text{q}}$ are left undetermined by the field equations, and that this corresponds to the freedom to add a $g$-mode to the interior solution.} Similarly, the angular components of the vorticity conservation condition fail to produce a definite assignment for $v^{\text{d}}$ and $v^{\text{o}}$, and the remaining $tr$ component is redundant.

I remark that none of the fluid variables determined by Eq.~\eqref{vortcons} depend explicitly on time, in contrast to the response to a gravitomagnetic tidal field (see Sec.~III of Paper II), which was characterized by fluid variables with a linear time dependence. Accordingly, I switch off the time dependence of the ansatz. The stationary ansatz automatically satisfies the relativistic Euler equation and, in the following section, I find that a completely stationary solution is compatible with the gravitoelectric sector's field equations.

\section{Field Equations}
\label{sec:efe}

\subsection{Zero-frequency modes}
\label{subsec:zerofreq}

Before going on to examine the field equations which govern the interior solution, I pause to explain the meaning of the freedom that remains in the fluid variables after the preceding section's analysis. The variables $\lbrace v^{\text{d}}, \hat{v}^{\text{q}}, \hat{v}^{\text{q}}_r, v^{\text{o}} \rbrace$ were left undetermined by the vorticity conservation condition, Eq.~\eqref{vortcons}; in this section, I find that the residual freedom in these variables is not eliminated by the Einstein field equations. Rather, as made clear by \citex{Lockitch_RotModesAnal}, it is associated with the freedom to add zero-frequency $r$- and $g$-modes to the interior solution.

The $g$-modes are polar (even-parity), stationary fluid disturbances which characterize the perturbation of a static, spherically symmetric material body \cite{Thorne_Pulsations}. To investigate how they manifest themselves in the context of this work, I set $\Omega = 0$, switch off the external tidal field, and focus on stationary perturbations in this discussion. I also suppress the multipole labels (e.g. $\text{d}$ and $\text{o}$) to indicate that the discussion is valid for all multipole orders $\ell$. A polar perturbation is described by the metric variables $\lbrace \hat{e}_{tt}, \hat{e}_{tr}, \hat{e}_{rr}, \hat{e} \rbrace$ and the fluid variables $\lbrace \hat{v}_r, \hat{v}, \hat{p} \rbrace$. By virtue of the field equations, the variables decouple into the groups $\lbrace \hat{e}_{tt}, \hat{e}_{rr}, \hat{e}, \hat{p} \rbrace$ and $\lbrace \hat{e}_{tr}, \hat{v}_r, \hat{v} \rbrace$. The first group of variables vanishes for a homogeneous perturbation (one not driven by an external tidal field). The second group, however, admits an infinite number of solutions when the fluid is barotropic; each one is characterized by a freely specifiable $\hat{v}_r$. These solutions define the class of zero-frequency $g$-modes. A free $g$-mode can be eliminated by making the assignment $\hat{v}_r = e^{-\psi}\hat{e}_{tr}$, which sets the corresponding component of the (contravariant) Eulerian velocity perturbation to zero \cite{Lockitch_RotModesAnal}.

The $r$-modes are axial (odd-parity), stationary disturbances of the fluid in a perturbed static, spherically symmetric material body \cite{Lockitch_RotModesAnal}. An axial perturbation is described by the metric variables $\lbrace f_t, f_r, v \rbrace$, where I have again omitted the multipole labels (e.g. $\text{q}$). In this case, the field equations admit another infinite set of solutions; each one is characterized by a freely specifiable $v$ and a vanishing $f_r$. These solutions define the class of zero-frequency $r$-modes, which are not restricted to barotropes. A free $r$-mode can be removed by setting $v = e^{-\psi} f_t$, which eliminates its associated (contravariant) Eulerian velocity perturbation \cite{Lockitch_RotModesAnal}.

To simplify the solution to the field equations, I choose to discard the free $r$- and $g$-modes when they appear. Nonetheless, the freedom to add zero-frequency modes to the solution remains, and the $r$- and $g$-modes can be restored at will. To illustrate the elimination of the zero-frequency modes concretely, I revisit the tidal perturbation introduced in Sec.~\ref{sec:spacetime}, and add a component $p_{tr}^{\text{tidal}} = e^{\text{q}}_{tr}(r) \mathcal{E}^{\text{q}}$ to the ansatz. The field equations imply that $e^{\text{q}}_{tr}$ is a solution to

\begin{equation}
 \left[8\pi r^2 (\mu+ p) - 3 \right] e^{\text{q}}_{tr} - 8\pi r^2 (\mu+p) e^{\psi} v^{\text{q}}_r = 0 . \label{gmodeefe}
\end{equation}
Since Eq.~\eqref{vqr} relates $v^{\text{q}}$ to $v^{\text{q}}_r$, and $v^{\text{q}}_r$ is otherwise unconstrained by the field equations, it is clear that the group of variables $\lbrace e^{\text{q}}_{tr}, v^{\text{q}}_r, v^{\text{q}} \rbrace$ represents a $g$-mode. The $g$-mode is removed by setting $v^{\text{q}}_r = e^{-\psi} e^{\text{q}}_{tr}$. Equation \eqref{gmodeefe} then produces the assignment $e^{\text{q}}_{tr} = 0$, and Eq.~\eqref{vqr} in turn sets $v^{\text{q}} = 0$.\footnote{To be more precise, Eq.~\eqref{vqr} implies that $\frac{d}{dr}(e^{-\psi} v^q) = 0$. The quadrupole piece of Eq.~\eqref{deltauA} is therefore $\delta u_A \sim [\text{constant}] \times e^{\psi} \mathcal{E}^{\text{q}}_A$, such that $\delta u^A \sim [\text{constant}]/r^2 \times e^{\psi} \Omega^{AB} \mathcal{E}^{\text{q}}_B$. Regularity at $r=0$ then demands that the constant vanishes, and hence that $v^{\text{q}} = 0$.} This choice explains the absence of a $p_{tr}^{\text{tidal}}$ component in the ansatz of Sec.~\ref{sec:spacetime}; for consistency with the metric ansatz, I must then set $v^{\text{q}}_r = 0 = v^{\text{q}}$.

\subsection{Field equations: $\ell = 2$}
\label{subsec:quad}

The metric ansatz of Sec.~\ref{sec:spacetime} and the fluid variables of Sec.~\ref{sec:fluid} are inserted into the Einstein field equations,

\begin{equation} \label{efe}
G_{\alpha\beta} = 8 \pi T_{\alpha\beta} \; ,
\end{equation}
which are then expanded to first order in $\Omega$ and $\mathcal{E}^{\text{q}}_{\mathsf{m}}$. Each component of Eq.~\eqref{efe} is decomposed in (scalar, vector and tensor) spherical harmonics. The field equations decouple according to multipole order, and I start by examining the $\ell = 2$ sector associated with the metric variables $\lbrace \hat{e}_{tt}^{\text{q}}, \hat{e}_{tr}^{\text{q}}, \hat{e}_{rr}^{\text{q}}, \hat{e}^{\text{q}} \rbrace$ and the fluid variables $\lbrace \hat{v}_r^{\text{q}}, \hat{v}^{\text{q}}, \hat{p}^{\text{q}} \rbrace$. The variables decouple into the groups $\lbrace \hat{e}_{tt}^{\text{q}}, \hat{e}_{rr}^{\text{q}}, \hat{e}^{\text{q}}, \hat{p}^{\text{q}} \rbrace$ and $\lbrace \hat{e}_{tr}^{\text{q}}, \hat{v}_r^{\text{q}}, \hat{v}^{\text{q}} \rbrace$.

The pressure perturbation $\hat{p}^{\text{q}}$ is eliminated with Eq.~\eqref{phatq}. The angular components of the field equations, together with the combined $rA$ and $rr$ components, further eliminate $\hat{e}^{\text{q}}_{rr}$ and $\hat{e}^{\text{q}}$. The sole remaining first-group variable, $\hat{e}^{\text{q}}_{tt}$, satisfies a homogeneous differential equation supplied by the $tt$ component of the field equations. This is the type of equation which governs a stationary, homogeneous perturbation of a static, spherically symmetric body; therefore, the first-group variables all vanish in accordance with the previous subsection's discussion.

The $tr$ component of the field equations gives rise to

\begin{equation} \label{ehatqtr}
\hat{e}^{\text{q}}_{tr} = \frac{e^{\psi }}{8 \pi  r^2 (\mu + p) -3 }  \left[8 \pi  r^2 (\mu + p) \hat{v}_r^{\text{q}}+\frac{1}{2} r^2 e^{-3 \psi } (1-\omega ) \frac{d e^{\text{q}}_{tt}}{dr}+\frac{1}{2} r e^{-3 \psi } \left(r \frac{d\omega}{dr} + 2 \omega- 2\right) e^{\text{q}}_{tt} \right] ,
\end{equation}
an algebraic equation for $\hat{e}^{\text{q}}_{tr}$, while the $tA$ components produce

\begin{eqnarray}
-8  \pi  r^2 (\mu + p) e^{\psi} \hat{v}^{\text{q}} = r^2 &f& \frac{d \hat{e}^{\text{q}}_{tr}}{dr} + \left[2 m-4 \pi  r^3 (\mu - p)\right] \hat{e}^{\text{q}}_{tr} + \frac{1}{3} r^3 e^{-2 \psi} \left(\frac{3m}{r}-r+4 \pi  r^2 p\right) (1-\omega ) \frac{d e^{\text{q}}_{tt}}{dr} \nonumber \\ + &&\frac{2}{3} r^2 e^{-2 \psi} \left\lbrace \left(1 + \frac{m}{r}\right) (1-\omega )+ \pi  r^2 (\mu + p) \left[1+\frac{d\mu}{dp}+ \left(3+\frac{d\mu}{dp}\right) \omega  \right]\right\rbrace e_{tt}^{\text{q}} .
\end{eqnarray}
The field equations do not fully determine the second-group variables $\lbrace \hat{e}_{tr}^{\text{q}}, \hat{v}_r^{\text{q}}, \hat{v}^{\text{q}} \rbrace$, and the residual freedom is interpreted as the freedom to add a $g$-mode to the solution. To simplify the description, I remove the $g$-mode by setting $\hat{v}^{\text{q}}_r = e^{-\psi} \hat{e}^{\text{q}}_{tr}$. Equation \eqref{ehatqtr} then reduces to

\begin{equation}
\hat{e}^{\text{q}}_{tr} = -\frac{1}{6} r e^{-2 \psi } \left[ (1-\omega ) r \frac{d e^{\text{q}}_{tt}}{dr} + \left(r \frac{d\omega}{dr} + 2 \omega - 2\right) e^{\text{q}}_{tt} \right] .
\end{equation}

The interior metric variables must be matched with the exterior solutions listed in Table IV of Paper I at $r=R$. It is straightforward to verify that the expression given for $\hat{e}^{\text{q}}_{tr}$ in Eq.~\eqref{ehatqtr} automatically agrees with its external counterpart at the body's surface regardless of the choice of $\hat{v}^{\text{q}}_r$, since $\mu$ and $p$ vanish at $r=R$ like $\mu \propto (1-r/R)^n$ and $p \propto (1-r/R)^{n+1}$, with $n >0$. The matching of $\hat{e}^{\text{q}}_{tt} = 0$ to its external version (subject to the change of notation described in footnote \ref{chiomega})

\begin{equation}
\hat{e}^{\text{q}}_{tt}[\text{ext}]  = \frac{8 I}{x^3}\left[-30x^3(x-1)^2 \ln{(1-1/x)} - \frac{5}{2} x (2x-1)(6x^2-6x-1) \right]\left(\mathfrak{E}^{\text{q}}-\frac{1}{120}\right) ,
\end{equation}
where $x:=r/2M$, determines the rotational-tidal Love number $\mathfrak{E}^{\text{q}}$ which appears in the exterior metric. The matching conditions produce the assignment $\mathfrak{E}^{\text{q}} = 1/120$. This result is independent of the material body's equation of state.

\subsection{Field equations: $\ell = 1$}
\label{subsec:dip}

In this subsection, I turn my attention to the $\ell = 1$ sector of the perturbation associated with the variables $\lbrace f^{\text{d}}_t, v^{\text{d}} \rbrace$. The metric variable $f^{\text{d}}_r$ does not figure in the ansatz because of the gauge choice $f^{\text{d}}_r = 0$ employed in Sec.~\ref{sec:spacetime}. The $tA$ components of the field equations give rise to the inhomogeneous differential equation

\begin{eqnarray}
 0 = r^2 &&f \frac{d^2 f^{\text{d}}_t}{dr^2} -4 \pi  r^3 \left(\mu+p\right) \frac{d f^{\text{d}}_t}{dr} + \left[ \frac{4 m}{r}- 2 + 8 \pi  r^2 \left(\mu+p\right) \right] f^{\text{d}}_t - 16 \pi r^2 (\mu+p) e^{\psi} v^{\text{d}} \nonumber \\ + &&\frac{2}{5} r^3 f e^{-2 \psi} \left(1 - \omega \right) \frac{d e^{\text{q}}_{tt}}{dr} + \frac{1}{5} r^2 e^{-2 \psi} \left[6(1-\omega) -4 \pi  r^2 (\mu+p) \left(1+\frac{d\mu}{dp}\right) \omega -4 \pi  r^2 (\mu+ p) \left(3+\frac{d\mu}{dp}\right) \right] e^{\text{q}}_{tt} \label{fdt}
\end{eqnarray}
for $f^{\text{d}}_t$. The fluid variable $v^{\text{d}}$ is left undetermined, and the corresponding residual freedom in the variables $\lbrace f^{\text{d}}_t, \linebreak f^{\text{d}}_r = 0, v^{\text{d}} \rbrace$ represents an $r$-mode. I choose to eliminate the $r$-mode to simplify the solution by setting $v^{\text{d}} = e^{- \psi} f^{\text{d}}_t$.

The matching conditions on the interior and exterior solutions require $f^{\text{d}}_t$ to agree at $r=R$ with the value appearing in Table IV of Paper I. In addition, it must satisfy regularity conditions at $r=0$. These boundary conditions fully determine $f^{\text{d}}_t$ inside and outside the body, including the residual gauge constant $\gamma^{\text{d}}$ which appears in the exterior solution. The precise value of $\gamma^{\text{d}}$ depends on the choice made for $v^{\text{d}}$.

\subsection{Field equations: $\ell = 3$}
\label{subsec:oct}

Finally, I examine the $\ell = 3$ sector of the perturbation associated with the variables $\lbrace f^{\text{o}}_t, f^{\text{o}}_r, v^{\text{o}} \rbrace$. The $rA$ components of the field equations indicate that $f^{\text{o}}_r = 0$. The metric variable $f^{\text{o}}_t$ is governed by the inhomogeneous differential equation

\begin{eqnarray}
0 =  r^2 & & f \frac{d^2 f^{\text{o}}_t}{dr^2} -4 \pi  r^3 \left(\mu+p\right) \frac{d f^{\text{o}}_t}{dr} + 4 \left[\frac{m}{r}-3 + 2 \pi  r^2 \left(\mu+p\right)\right] f^{\text{o}}_t \nonumber \\ - && 16\pi r^2 (\mu+p) e^{\psi} v^{\text{o}} \nonumber + r^3 e^{-2 \psi } (1-\omega) \left(\frac{9 m}{r}-2 + 20 \pi  r^2 p \right) \frac{d e^{\text{q}}_{tt}}{dr} \nonumber \\ + && 2 r^2 e^{-2 \psi} \left[ \left(\frac{5m}{r}+2\right)(1-\omega) + 2\pi  r^2 (\mu+p) \left(6+\frac{d\mu}{dp}\right) \omega +  2 \pi  r^2 (\mu+p) \left(\frac{d\mu}{dp}-2\right) \right] e^{\text{q}}_{tt} \label{fot}
\end{eqnarray}
which results from the $tA$ components of the field equations. The fluid variable $v^{\text{o}}$ is not specified by the field equations, and the associated freedom in the variables $\lbrace f^{\text{o}}_t , f^{\text{o}}_r = 0, v^{\text{o}} \rbrace$ represents an $r$-mode. I choose again, for simplicity, to eliminate this freedom in the solution; the assignment $v^{\text{o}} = e^{- \psi} f^{\text{o}}_t$ has the desired effect.

The matching conditions on the interior and exterior solutions demand that $f^{\text{o}}_t$ agree at $r=R$ with the value

\begin{eqnarray}
 (8 I M x^3)^{-1} f^{\text{o}}_t[\text{ext}]  &=& -\frac{10}{3 x^6} \left[ - \frac{3}{2}x^2(5x-2)\ln{(1-1/x)}-\frac{3}{4}x(10x+1) \right] K_2^{\text{el}} \nonumber \\ &&+ \frac{2}{x^6} \left[ 210x^5(3x-2)(x-1)\ln{(1-1/x)} + \frac{7}{2} x^2(180x^4-210x^3+30x^2+5x+1) \right] \mathfrak{F}^{\text{o}} \nonumber \\ &&-\frac{5}{12x^3}+\frac{1}{6x^4} \label{fotout}
\end{eqnarray}
from Table IV of Paper I. It must moreover satisfy regularity conditions at $r=0$. These boundary conditions fully specify $f^{\text{o}}_t$ inside and outside the body, thereby determining the rotational-tidal Love number $\mathfrak{F}^{\text{o}}$ which appears in the exterior solution.

Unlike $\mathfrak{E}^{\text{q}}$, the rotational-tidal Love number $\mathfrak{F}^{\text{o}}$ depends sensitively on the body's equation of state. In order to present quantitative results for $\mathfrak{F}^{\text{o}}$, I specialize to the polytropic equation of state Eq.~\eqref{mpoly}. The explicit boundary conditions for integration of Eq.~\eqref{fot} can be worked out based on the asymptotic behavior of the fluid variables as $r \rightarrow 0$: they approach their central values like $(p-p_c) \propto r^{2}$, $(\mu-\mu_c) \propto r^{2}$, $m \propto r^3$ and $(\psi - \psi_c) \propto r^2$. The integration of the differential equation is performed numerically from $r=0$ to $r=R$ for various choices of the parameters $(n, M/R)$ which specify the polytropic model. The results remain finite in the limit of zero compactness when scaled with a factor of $(M/R)^5$. Accordingly, I define $\mathfrak{f}^{\text{o}} := -(2M/R)^5 \mathfrak{F}^{\text{o}}$ as the scale-free version of the rotational-tidal Love number $\mathfrak{F}^{\text{o}}$, and plot it as a function of the polytrope's compactness in Fig.~\ref{fig:fo}.

I note that the value of $\mathfrak{F}^{\text{o}}$ depends on the choice made for $v^{\text{o}}$; it is sensitive to the presence of internal $r$-modes. The numerical results I display in Fig.~\ref{fig:fo} refer to a material body that is free of $r$-modes, and would be modified if such a mode were incorporated in the solution.

\section{Gravitomagnetic Tidal Response}
\label{sec:gravitomag}

In this section, I return to the scenario envisioned in Paper II: I replace the gravitoelectric tidal field generated by $\mathcal{E}_{ab}$ with a gravitomagnetic one generated by $\mathcal{B}_{ab}$. The metric ansatz and fluid variables of Secs. II and III of Paper II replace those introduced in this paper. I revisit the interior solution presented in Sec. IV of Paper II with the goal of completing the integration of the Einstein field equations and computing the rotational-tidal Love numbers associated with the gravitomagnetic sector of the tidal response. I also calculate the velocity perturbation induced by the gravitomagnetic field, and estimate its size for a neutron star in a binary system.

The manipulations carried out in this section rely heavily on the developments presented in Paper II. I reproduce only the most relevant equations here; the reader is referred to Paper II for further details. I shall focus primarily on the $tt$, $tr$ and $tA$ components of the metric perturbation, as well as the Eulerian velocity perturbation. The bilinear perturbations

\begin{subequations}
\begin{eqnarray}
p_{tt}^{\text{bilinear}} &=& k^{\text{d}}_{tt}(r) \mathcal{K}^{\text{d}} + k^{\text{o}}_{tt}(r) \mathcal{K}^{\text{o}} , \\
p_{tr}^{\text{bilinear}} &=& k^{\text{d}}_{tr}(t,r) \mathcal{K}^{\text{d}} + k^{\text{o}}_{tr}(t,r) \mathcal{K}^{\text{o}}
\end{eqnarray}
\end{subequations}
make up the $tt$ and $tr$ components of the metric perturbation. The $tA$ components consist of the tidal perturbation

\begin{equation}
p_{tA}^{\text{tidal}} = b^{\text{q}}_t(r) \mathcal{B}^{\text{q}}_A
\end{equation}
and the bilinear perturbation

\begin{equation}
p_{tA}^{\text{bilinear}} = \hat{b}^{\text{q}}_{t}(t,r) \hat{\mathcal{B}}^{\text{q}}_A .
\end{equation}
The tidal potentials $\lbrace \mathcal{K}^{\text{d}}, \mathcal{K}^{\text{o}}, ..., \hat{\mathcal{B}}^{\text{q}}_A \rbrace$ are defined in Sec. II of Paper I.\footnote{Recall that I construct the tidal moments with $\Omega$ instead of $\chi$ throughout this paper. The tidal moments here therefore differ by a factor of $\chi/\Omega = I/M^2$ from those of Paper I.} The radial function $b^{\text{q}}_t$ is governed by the field equation

\begin{equation} \label{bqt}
r^2 f \frac{d^2 b^{\text{q}}_t}{d r^2} - 4\pi r^3 (\mu + p) \frac{d b^{\text{q}}_t}{dr} - 2 \left[ 3-\frac{2m}{r} - 4\pi r^2 (\mu + p) \right] b^{\text{q}}_t = 0 ,
\end{equation}
and it agrees with its external counterpart

\begin{equation} \label{bqtR}
b^{\text{q}}_t[\text{ext}] = \frac{16 M^3 x^3}{3} \left\lbrace 1 - \frac{1}{x} - \frac{3}{x^5} \left[ 20 x^4 (x-1) \ln{(1-1/x)} + \frac{5}{3} x (12 x^3 - 6x^2 - 2x - 1) \right] K_2^{\text{mag}} \right \rbrace
\end{equation}
from Table IV of Paper I at $r=R$. The Eulerian velocity perturbation is decomposed as

\begin{equation} \label{dumag}
\delta u_r = v^{\text{d}}_r (t,r) \mathcal{K}^{\text{d}} + v^{\text{o}}_r (t,r) \mathcal{K}^{\text{o}} , \qquad \delta u_A =  v^{\text{d}} (t,r) \mathcal{K}^{\text{d}}_A + \hat{v}^{\text{q}} (t,r) \hat{\mathcal{B}}^{\text{q}}_A + v^{\text{o}} (t,r) \mathcal{K}^{\text{o}}_A .
\end{equation}
The complete set of metric and fluid variables for the gravitomagnetic sector of the response, presented in Secs.~II and III of Paper II, is inserted into the Einstein field equations, which are expanded to first order in $\Omega$ and $\mathcal{B}^{\text{q}}_{\mathsf{m}}$. The resulting equations, first displayed in Sec.~IV of Paper II, are revisited below.

\subsection{Field equations: $\ell = 2$}
\label{subsec:magquad}

The $\ell = 2$ sector of the perturbation sourced by $\mathcal{B}_{ab}$ is characterized by the metric and fluid variables $\lbrace \hat{b}^{\text{q}}_t, \hat{b}^{\text{q}}_r, \hat{v}^{\text{q}} \rbrace$. The Einstein field equations imply that $\hat{b}^{\text{q}}_t$ is linear in time and satisfies

\begin{equation} \label{bhatqeq}
r^2 f \frac{d^2 \hat{b}^{\text{q}}_{t}}{dr^2} - 4\pi r^3 (\mu + p) \frac{d \hat{b}^{\text{q}}_t}{dr} - 2 \left[ 3 - \frac{2m}{r} -4\pi r^2 (\mu+p) \right] \hat{b}^{\text{q}}_t - 16 \pi r^2 (\mu + p) e^{\psi} \hat{v}^{\text{q}} = 0 .
\end{equation}
The statement of vorticity conservation implies that

\begin{equation} \label{vq}
v^{\text{q}} = e^{-\psi} \left( \partial_t \xi^{\text{q}} + b^{\text{q}}_t \right) = 0 ,
\end{equation}
and that

\begin{equation} \label{vhatq}
\hat{v}^{\text{q}} = -\frac{1}{3} e^{-\psi} \omega \xi^{\text{q}} .
\end{equation}
Because the exterior solution of Paper I is stationary, the matching conditions on the interior and exterior metric require that $\hat{b}^{\text{q}}_{t1} = 0 = d \hat{b}^{\text{q}}_{t1}/dr$ at $r=R$, with $\hat{b}^{\text{q}}_t:= t \hat{b}^{\text{q}}_{t1}$. The matching of $\hat{b}^{\text{q}}_t$ to its external expression

\begin{equation}
\hat{b}^{\text{q}}_{t}[\text{ext}]  = \frac{16 I M}{x^2}\left[20x^4(x-1) \ln{(1-1/x)} + \frac{5}{3} x (12x^3-6x^2-2x-1) \right]\left(\mathfrak{B}^{\text{q}}-\frac{1}{120}\right) .
\end{equation}
then produces the assignment $\mathfrak{B}^{\text{q}} = 1/120$ for the rotational-tidal Love number appearing therein. This result is universal for all material bodies in the absence of internal $r$-modes. The addition of an $r$-mode characterized by $\lbrace \hat{b}^{\text{q}}_{t0}(r), \hat{b}^{\text{q}}_{r0}(r) = 0, \hat{v}^{\text{q}}_{0}(r) \rbrace$ would impact the value of $\mathfrak{B}^{\text{q}}$ through the change in the value of $\hat{b}^{\text{q}}_t := t \hat{b}^{\text{q}}_{t1} + \hat{b}^{\text{q}}_{t0}$ and its first derivative at $r=R$.

\begin{figure}
\includegraphics[width=\textwidth]{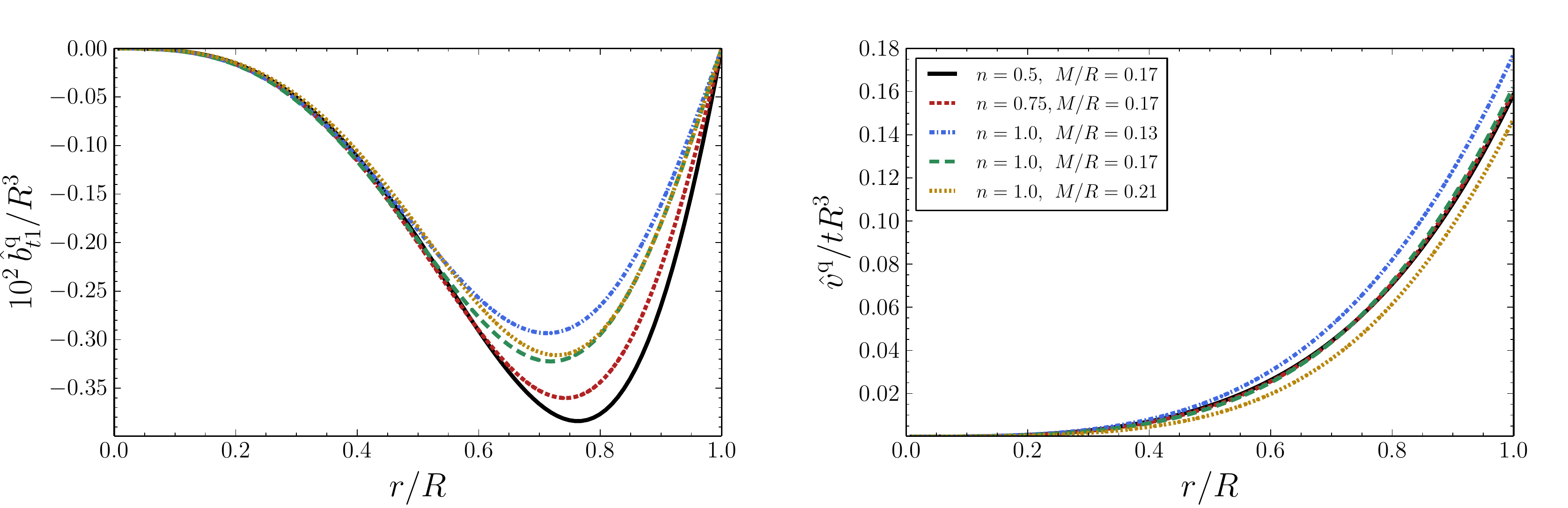}
\caption{\label{fig:velq} \label{fig:bhatq} Numerical solution for the radial function $\hat{b}^{\text{q}}_{t1}$ (left panel) and radial profile of the fluid variable $\hat{v}^{\text{q}}$ (right panel) for polytropes of index $n$ and compactness $M/R$.}
\end{figure}

To calculate the fluid variable $\hat{v}^{\text{q}}$ on the basis of Eqs.~\eqref{vq} and \eqref{vhatq}, Eq.~\eqref{bqt} must be integrated from $r=0$ to $r=R$. This requires specifying an equation of state for the body, and I select the relation Eq.~\eqref{mpoly} for mass polytropes as above. The resulting radial profile for $\hat{v}^{\text{q}}$ is plotted for various polytropic models in Fig.~\ref{fig:velq}. I have also integrated Eq.~\eqref{bhatqeq} and plotted the function $\hat{b}^{\text{q}}_{t1}$ in the same figure.

\subsection{Field equations: $\ell = 1$}
\label{subsec:magdip}

The $\ell = 1$ sector of the perturbation involves the metric variables $\lbrace k^{\text{d}}_{tt}, k^{\text{d}}_{tr}, k^{\text{d}}_{rr} \rbrace$ and the fluid variables $\lbrace v^{\text{d}}_r, v^{\text{d}}, p^{\text{d}} \rbrace$, which decouple into the groups $\lbrace k^{\text{d}}_{tt}, k^{\text{d}}_{rr}, p^{\text{d}} \rbrace$ and $\lbrace k^{\text{d}}_{tr}, v^{\text{d}}_r, v^{\text{d}} \rbrace$. The field equations impart a linear time dependence to the metric variable $k^{\text{d}}_{tr} := t k^{\text{d}}_{tr1}$, and produce the assignments

\begin{subequations} \label{dipolev}
\begin{eqnarray}
v^{\text{d}}_r &=& - t \frac{e^{-\psi} \left[ 1 - 8 \pi r^2 (\mu + p) \right]}{8 \pi r^2 (\mu + p)} k^{\text{d}}_{tr1} , \\
v^{\text{d}} &=& - t \frac{e^{-\psi}}{16 \pi r^2 (\mu + p)} \left\lbrace r^2 f \frac{d k^{\text{d}}_{tr1}}{dr} + 2 \left[ m - 2\pi r^3 (\mu - p) \right] k^{\text{d}}_{tr1} \right\rbrace .
\end{eqnarray}
\end{subequations}
The function $k^{\text{d}}_{tr1}$ satisfies the inhomogeneous differential equation

\begin{eqnarray} \label{kdtr1}
0 = \,&r^2& f \frac{d^2 k^{\text{d}}_{tr1}}{dr^2} + \left[ 3(m-4\pi r^3 \mu) + (m + 4\pi r^3 p) \frac{d\mu }{dp} \right] \frac{d k^{\text{d}}_{tr1}}{dr} \nonumber \\ &-& \frac{2}{r^2 f} \left\lbrace \left[ 1 - 10 \pi r^2 (\mu +p) + 16 \pi^2 r^4 p^2 \right] r^2 + 4\pi r^3 (5\mu + 7p)m - 3m^2 - (m+4\pi r^3 p)^2 \frac{d\mu}{dp} \right\rbrace k^{\text{d}}_{tr1} \nonumber \\ &-& \frac{48 \pi}{5} r^2 (\mu+p)\omega \frac{d b^{\text{q}}_t}{dr} - \frac{96 \pi}{5}(\mu + p)\left( r^2 \frac{d\omega}{dr} + \frac{r - 4m - 8\pi r^3 p}{f} \omega \right) b^{\text{q}}_t .
\end{eqnarray}
Since $k^{\text{d}}_{tr}$ vanishes in the exterior solution of Paper I, the matching conditions require that $ k^{\text{d}}_{tr1} = 0 = d k^{\text{d}}_{tr1}/dr $ at $r=R$.

\begin{figure}
\includegraphics[width=0.75\textwidth]{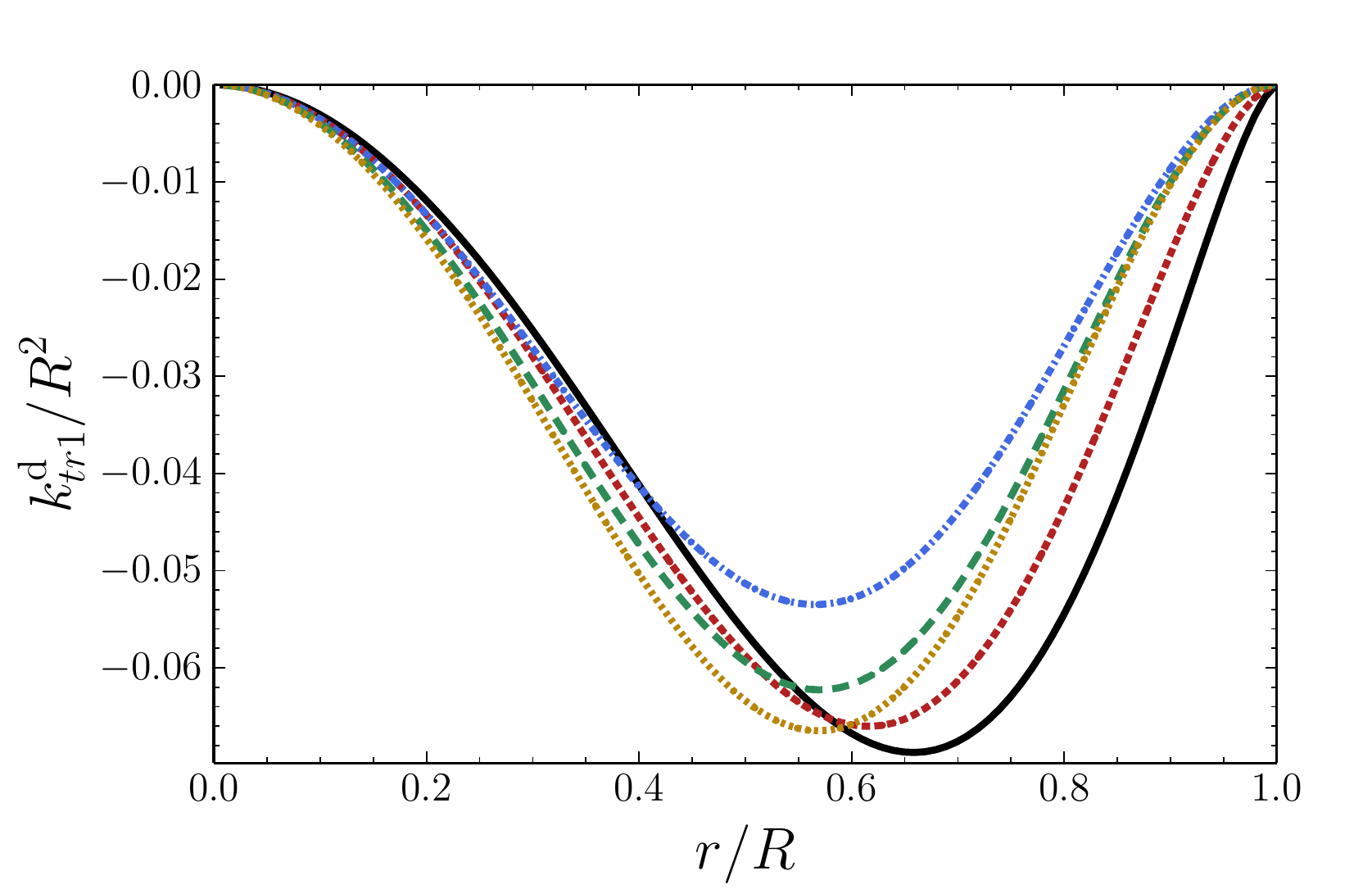}
\caption{\label{fig:kdtr1} Numerical solution for the radial function $k^{\text{d}}_{tr1}$ for polytropes of index $n$ and compactness $M/R$.}
\end{figure}

\begin{figure}
\includegraphics[width=\textwidth]{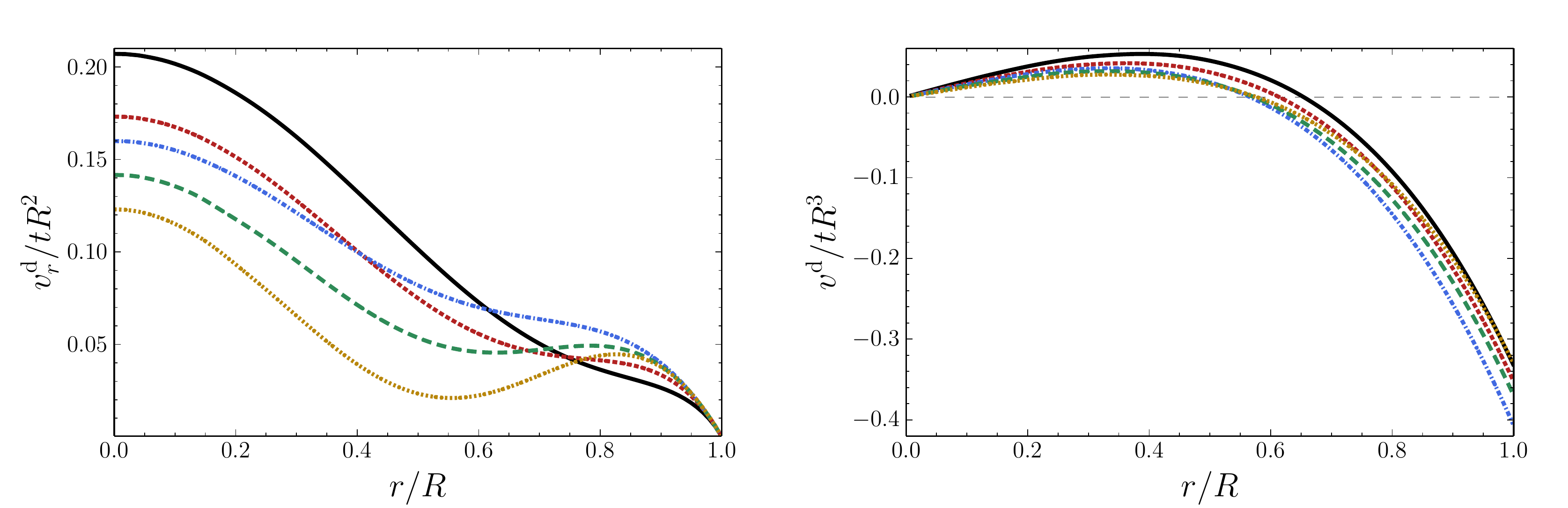}
\caption{\label{fig:veld} Radial profiles of the fluid variables $v^{\text{d}}_r$ (left panel) and $v^{\text{d}}$ (right panel) for polytropes of index $n$ and \mbox{compactness $M/R$}.}
\end{figure}

To calculate $v^{\text{d}}_r$ and $v^{\text{d}}$ via Eq.~\eqref{dipolev}, I integrate Eq.~\eqref{kdtr1} from $r=0$ to $r=R$. The numerical solution $k^{\text{d}}_{tr1}$, and the radial profiles of the fluid variables, are shown in Figs.~\ref{fig:kdtr1} and \ref{fig:veld}, respectively, for different mass polytropes. In addition, I note that the matching conditions on $k^{\text{d}}_{tt}$, which is determined by the field equation

\begin{eqnarray} \label{kdtteq}
0 &=& r \frac{d k^{\text{d}}_{tt}}{dr} + \frac{2 \left[ m -2\pi r^3(\mu+p) \right](r-m+4\pi r^3 p)}{rf(m+4\pi r^3 p)} k^{\text{d}}_{tt} - \frac{r^2 (r-m+4\pi r^3 p)}{2(m+4\pi r^3 p)} \frac{d k^{\text{d}}_{tr1}}{dr} \nonumber \\ &&+ \frac{\left[ 1 + 2\pi r^2 (1+ 4\pi r^2 p)(\mu - p) \right]r^2 -\left[ 5+ 2\pi r^2 (\mu +p) \right]rm+ 5m^2}{f(m+4\pi r^3 p)} k^{\text{d}}_{tr1} - \frac{3r}{10} \left[ \frac{r(r-m+4\pi r^3 p)}{m+4\pi r^3 p} \frac{d\omega}{dr} - 4\omega + 4 \right] \frac{d b^{\text{q}}_t}{dr} \nonumber \\ &&- \frac{3(r-m+4\pi r^3 p)}{5(m+4\pi r^3 p)} \left[ r \frac{d\omega}{dr} - 4 \frac{m+ 2\pi r^3 (\mu+p)}{rf}\omega + 4 \frac{m- 2\pi r^3(\mu+p)}{rf} \right] b^{\text{q}}_t ,
\end{eqnarray}
set the value of the residual gauge constant $c^d$ appearing in Table I of Paper I. The matching is unaffected by the addition of a $g$-mode $\lbrace k^{\text{d}}_{tr0}(r), v^{\text{d}}_{r0}(r), v^{\text{d}}_0(r) \rbrace$ to the solution, since only the time derivative of $k^{\text{d}}_{tr} := t k^{\text{d}}_{tr1} + k^{\text{d}}_{tr0}$ appears in Eq.~\eqref{kdtteq}.

\subsection{Field equations: $\ell = 3$}
\label{subsec:magoct}

The metric variables $\lbrace k^{\text{o}}_{tt}, k^{\text{o}}_{tr}, k^{\text{o}}_{rr}, k^{\text{o}} \rbrace$ and fluid variables $\lbrace v^{\text{o}}_r, v^{\text{o}}, p^{\text{o}} \rbrace$ characterize the $\ell = 3$ sector of the perturbation. By virtue of the Einstein field equations, they decouple into the groups $\lbrace k^{\text{o}}_{tt}, k^{\text{o}}_{rr}, k^{\text{o}}, p^{\text{o}} \rbrace$ and $\lbrace k^{\text{o}}_{tr}, v^{\text{o}}_r, v^{\text{o}} \rbrace$. The field equations further determine that $k^{\text{o}}_{tr} := t k^{\text{o}}_{tr1}$ is linear in time, and yield the assignments

\begin{subequations} \label{octupolev}
\begin{equation}
v^{\text{o}}_r = - t \frac{e^{-\psi} \left[ 3 - 4 \pi r^2 (\mu + p) \right]}{4 \pi r^2 (\mu + p)} k^{\text{o}}_{tr1} ,
\end{equation}
\begin{equation} \label{vo}
v^{\text{o}} = - t \frac{3 e^{-\psi}}{16 \pi r^2 (\mu + p)} \left\lbrace r^2 f \frac{d k^{\text{o}}_{tr1}}{dr} + 2 \left[ m - 2\pi r^3 (\mu - p) \right] k^{\text{o}}_{tr1} \right\rbrace .
\end{equation}
\end{subequations}
The function $k^{\text{o}}_{tr1}$ is a solution to the inhomogeneous differential equation

\begin{eqnarray} \label{kotr1}
0 = && r^2 f \frac{d^2 k^{\text{o}}_{tr1}}{dr^2} + \left[ 3(m-4\pi r^3 \mu) + (m + 4\pi r^3 p) \frac{d\mu }{dp} \right] \frac{d k^{\text{o}}_{tr1}}{dr} \nonumber \\ &-& \frac{2}{r^2 f} \left\lbrace 2 \left[ 3 - 5 \pi r^2 (\mu +p) + 8 \pi^2 r^4 p^2 \right] r^2 - 2 \left[ 5 - 2\pi r^2 (5\mu + 7p) \right]r m - 3m^2 - (m+4\pi r^3 p)^2 \frac{d\mu}{dp} \right\rbrace k^{\text{o}}_{tr1} \nonumber \\ &-& \frac{32 \pi}{3} r^2 (\mu+p)\omega \frac{d b^{\text{q}}_t}{dr} + \frac{16 \pi}{3}(\mu + p)\left[ r^2 \frac{d\omega}{dr} + 2 \frac{3r - 7m - 4\pi r^3 p}{f} \omega \right] b^{\text{q}}_t .
\end{eqnarray}
The fact that $k^{\text{o}}_{tr}$ vanishes in the exterior solution of Paper I means that the matching conditions on the function are $ k^{\text{o}}_{tr1} = 0 = d k^{\text{o}}_{tr1}/dr $ at $r=R$.

\begin{figure}
\includegraphics[width=\textwidth]{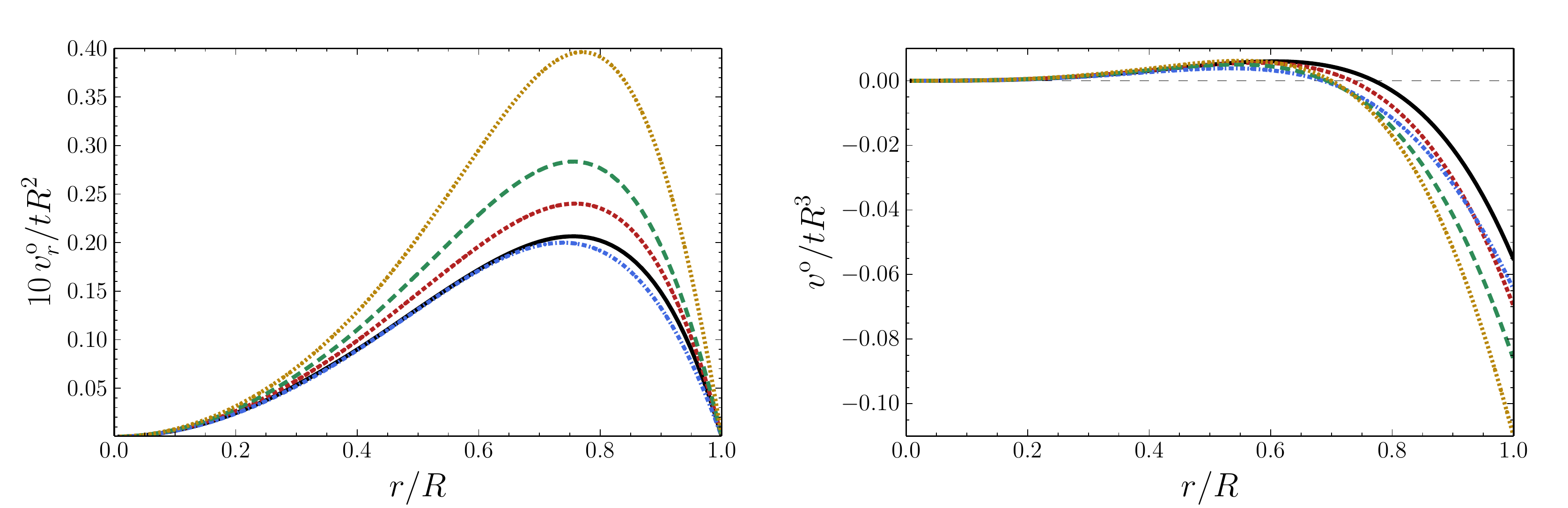}
\caption{\label{fig:velo} Radial profiles of the fluid variables $v^{\text{o}}_r$ (left panel) and $v^{\text{o}}$ (right panel) for polytropes of index $n$ and \mbox{compactness $M/R$}.}
\end{figure}

\begin{figure}
\includegraphics[width=0.75\textwidth]{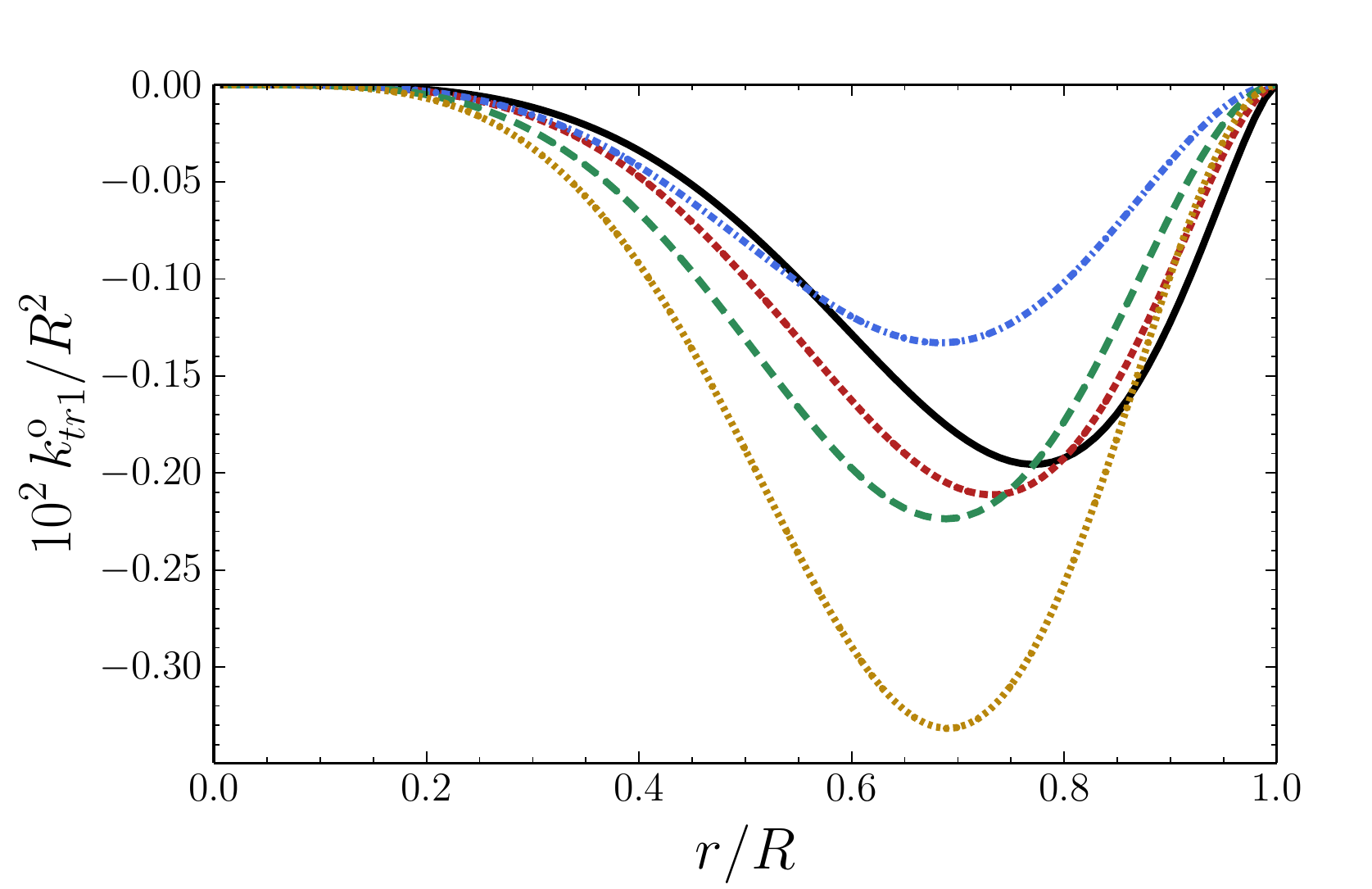}
\caption{\label{fig:kotr1} Numerical solution for the radial function $k^{\text{o}}_{tr1}$ for polytropes of index $n$ and compactness $M/R$.}
\end{figure}

I calculate $v^{\text{o}}_r$ and $v^{\text{o}}$ via Eq.~\eqref{octupolev} by integrating Eq.~\eqref{kotr1} from $r=0$ to $r=R$. The resulting radial profiles are plotted in Fig.~\ref{fig:velo} for various mass polytropes. The numerical solution for $k^{\text{o}}_{tr1}$ is shown in Fig.~\ref{fig:kotr1}. The field equations also produce the inhomogeneous differential equation

\begin{eqnarray} \label{kott}
0 &=& r^2 f \frac{d^2k^{\text{o}}_{tt}}{dr^2} +2 \left[1 -\frac{3 m}{r}-2 \pi  r^2 (\mu+3 p )\right] r \frac{dk^{\text{o}}_{tt}}{dr} + 4 \left[\pi  r^2 (\mu + p ) (3+\frac{d\mu}{dp}) - 3\right] k_{tt}^{\text{o}} \nonumber + \frac{1}{2} r^2 f \left(\frac{d\mu}{dp}-1 \right) \frac{dk^{\text{o}}_{tr1}}{dr} \nonumber \\ &&+\left\lbrace \left(11+\frac{d\mu}{dp}\right) m+ 2\pi  r^3 \left[ (\mu + 7p) - (\mu - p)\frac{d\mu}{dp}\right] -4r \right\rbrace k_{tr1}^{\text{o}} + S_1 r \frac{db^{\text{q}}_t}{dr} + S_0 b^{\text{q}}_t ,
\end{eqnarray}
obtained by combining Eqs.~(14.4a) and (14.4b) of Paper II, for the time-independent metric variable $k^{\text{o}}_{tt}$, where

\begin{subequations}
\begin{eqnarray}
S_1 &=& -\frac{2}{3} \left\lbrace\frac{5 m^2}{r^2} + 3-16 \pi ^2 r^4 p^2 +4 \pi  r^2 \mu -\frac{m}{r} \left[9 + 8 \pi  r^2 (\mu +p )\right] \right\rbrace r \frac{d\omega}{dr} \nonumber \\ &&-\frac{4}{3} \left\lbrace 3-\frac{m}{r} \left[9-4 \pi  r^2 (\mu + p )\right]-4 \pi  r^2 p \left[3-4 \pi  r^2 (\mu + p )\right]\right\rbrace \omega + 4 \left(1-\frac{3 m}{r}-4 \pi  r^2 p \right) , \\
S_0 &=& \frac{2}{3}\left\lbrace \frac{10m^2}{r^2}+4 \pi  r^2 \left[\left(3-8 \pi  r^2 p \right)p+2 \mu \right]-\frac{m}{r} \left[3+16 \pi  r^2 (\mu + p )\right]\right\rbrace r\frac{d\omega}{dr} \nonumber \\ &&+ \frac{4}{3} \left\lbrace\frac{m}{r} \left[6+8 \pi  r^2 (\mu + p )\right]+ \left[9-4 \pi  r^2 (\mu + p) \left(6+\frac{d\mu}{dp}-8 \pi  r^2 p \right) \right]\right\rbrace \omega \nonumber \\ &&- 4 \left\lbrace\frac{2 m}{r}- \left[2 \pi  r^2(\mu + p ) \left(1+\frac{d\mu}{dp}\right) - 3 \right]\right\rbrace .
\end{eqnarray}
\end{subequations}
At $r=R$, $k^{\text{o}}_{tt}$ matches its external version

\begin{eqnarray}
-(4Ix^2)^{-1} k^{\text{o}}_{tt}[\text{ext}] &=& \frac{1}{x^7} \left[ - 10 x^4(x-1)(280x^3-420 x^2+140x+3)\ln{(1-1/x)} \right] K_2^{\text{mag}} \nonumber \\ &&+ \frac{1}{x^7} \left[-2800x^7+5600x^6 - \frac{9100}{3}x^5 + \frac{610}{3}x^4 + \frac{115}{3}x^3 +5 x^2 - \frac{5}{6}x - \frac{5}{6} \right] K_2^{\text{mag}} \nonumber \\ &&+ \frac{2}{x^6} \left[ -420x^4(2x-1)(x-1)^2\ln{(1-1/x)} - 7x^2(120x^4-240x^3+130x^2-10x-1)\right] \mathfrak{K}^{\text{o}} \nonumber \\ &&+ \frac{1}{2x^2} - \frac{1}{2x^3}
\end{eqnarray}
from Table IV of Paper I; this condition, together with regularity at $r=0$, is sufficient to fully determine $k^{\text{o}}_{tt}$ inside and outside the body, including the rotational-tidal Love number $\mathfrak{K}^{\text{o}}$ which appears in the exterior solution. The presence of a $g$-mode within the body does not affect the value of $\mathfrak{K}^{\text{o}}$, since the addition of terms $\lbrace k^{\text{o}}_{tr0}(r), v^{\text{o}}_{r0}(r), v^{\text{o}}_0(r) \rbrace$ to the solution does not impact Eq.~\eqref{kott}, in which only the time derivative of $k^{\text{o}}_{tr} := t k^{\text{o}}_{tr1}+k^{\text{o}}_{tr0}$ appears.

Like $\mathfrak{F}^{\text{o}}$, the rotational-tidal Love number $\mathfrak{K}^{\text{o}}$ depends on the body's equation of state. Specializing to the polytropic form Eq.~\eqref{mpoly}, I integrate Eq.~\eqref{kott} numerically from $r=0$ to $r=R$ for several choices of the parameters $(n, M/R)$. The scale-free version of the rotational-tidal Love number, $\mathfrak{k}^{\text{o}} := -(2M/R)^5 \mathfrak{K}^{\text{o}}$, is plotted as a function of compactness $M/R$ in Fig.~\ref{fig:ko}. The scale-free nature of $\mathfrak{k}^{\text{o}}$ is ensured by the fact that it remains finite in the $M/R \rightarrow 0$ limit.

For a polytrope of index $n=1$, the rotational-tidal Love number has the value of $\mathfrak{k}^{\text{o}} = 9.0493 \times 10^{-2}$ in the zero-compactness limit. This figure agrees to one part in $10^{5}$ with the result of Poisson and Dou\c cot \cite{Poisson_PNDynResp}, who calculated the same quantity in a post-Newtonian approximation. The small discrepancy reflects the accuracy of my numerical integrations near $M/R = 0$.

\subsection{Tidal currents}
\label{subsec:velpert}

The external gravitomagnetic tidal field induces time-dependent velocity perturbations, despite the assumed time independence of $\mathcal{B}_{ab}$. As argued in Paper II, and shown convincingly by the post-Newtonian analysis of \citex{Poisson_PNDynResp}, the dynamical velocity perturbations represent time-varying internal currents which are driven by the zero-frequency modes of the fluid when the gravitomagnetic tidal field is allowed to couple to the body's spin. In this subsection, I calculate the amplitude of these currents at the equator  of a neutron star in a binary system of relevance to LIGO.

I suppose that the reference body is a neutron star in a binary system with a companion of mass $M'$ moving on a circular orbit of radius $b$ and orbital angular frequency $\Omega_{\text{orb}} = [(M+M')/b^3]^{1/2}$. I adopt a Cartesian coordinate system oriented so that the orbital plane lies in the $x$-$y$ plane. In these coordinates, the gravitomagnetic tidal quadrupole moment sourced by the companion's orbital motion has nonvanishing components

\begin{equation} \label{tidmom}
\mathcal{B}_{13} = - \frac{3M' \Omega_{\text{orb}}}{b^2} \cos{\Omega_{\text{orb}} t} , \qquad \mathcal{B}_{23} = - \frac{3M' \Omega_{\text{orb}}}{b^2} \sin{\Omega_{\text{orb}} t}
\end{equation}
to leading order in a post-Newtonian expansion \cite{Poisson_TidDefRotBH}.

In Paper II, and so far in this work, I have assumed that the orbital time scale $T_{\text{orb}} \sim 1/\Omega_{\text{orb}}$ is much longer than the internal hydrodynamical time scale $T_{\text{int}} \sim (R^3/M)^{1/2}$ for the body's tidal response, so that the tidal moments are effectively stationary. The assumption of strict stationarity is responsible for the linear time dependence of $\hat{v}^{\text{q}}$ which follows from Eqs.~\eqref{vq} and \eqref{vhatq}. I now relax this assumption to better model the binary setting, and allow the tidal moments to retain the sinusoidal time dependence exhibited in Eq.~\eqref{tidmom}.

The effect of taking $\mathcal{B}^{\text{q}}_{\mathsf{m}} = \mathcal{B}^{\text{q}}_{\mathsf{m}}(t)$ can be understood by examining Eq.~\eqref{vq}, in which I reinsert the implicit factors of the tidal potential $\mathcal{B}^{\text{q}}_A$:

\begin{equation}
e^{-\psi} \left[ \partial_t \left( \xi^{\text{q}} \mathcal{B}^{\text{q}}_A \right) + b^{\text{q}}_t \mathcal{B}^{\text{q}}_A \right] = 0 .
\end{equation}
Previously, for stationary $\mathcal{B}^{\text{q}}_{\mathsf{m}}$, this statement produced a linear time dependence $\xi^{\text{q}} \mathcal{B}^{\text{q}}_A = - t b^{\text{q}}_t \mathcal{B}^{\text{q}}_A$. In contrast, for time-dependent $\mathcal{B}^{\text{q}}_{\mathsf{m}}$, it yields a bounded time dependence

\begin{equation}
\xi^{\text{q}} \mathcal{B}^{\text{q}}_A = - b^{\text{q}}_t \int_0^t \mathcal{B}^{\text{q}}_A dt' = - b^{\text{q}}_t \sum_{\mathsf{m}} \int_0^t \mathcal{B}^{\text{q}}_{\mathsf{m}}(t') dt' X_A^{2 \mathsf{m}} .
\end{equation}
The net effect of the tidal moments' sinusoidal time dependence is therefore to take

\begin{equation}
t \mathcal{B}^{\text{q}}_{\mathsf{m}} \rightarrow \int_0^t \mathcal{B}^{\text{q}}_{\mathsf{m}}(t') dt' = \frac{1}{\Omega_{\text{orb}}} \int_0^{\Phi} \mathcal{B}^{\text{q}}_{\mathsf{m}}(\Phi') d\Phi' ,
\end{equation}
where $\Phi := \Omega_{\text{orb}} t$ is the orbital phase. Because the spherical-harmonic coefficients $\mathcal{K}^{\text{d}}_{\mathsf{m}}$, $\hat{\mathcal{B}}^{\text{q}}_{\mathsf{m}}$ and $\mathcal{K}^{\text{o}}_{\mathsf{m}}$ are related to $\mathcal{B}^{\text{q}}_{\mathsf{m}}$ (see Table I of Paper I), the time dependence further implies

\begin{equation}
t \mathcal{K}^{\text{d}}_{\mathsf{m}} \rightarrow \frac{1}{\Omega_{\text{orb}}} \int_0^{\Phi} \mathcal{K}^{\text{d}}_{\mathsf{m}}(\Phi') d\Phi' , \qquad t \hat{\mathcal{B}}^{\text{q}}_{\mathsf{m}} \rightarrow \frac{1}{\Omega_{\text{orb}}} \int_0^{\Phi} \hat{\mathcal{B}}^{\text{q}}_{\mathsf{m}}(\Phi') d\Phi' , \qquad t \mathcal{K}^{\text{o}}_{\mathsf{m}} \rightarrow \frac{1}{\Omega_{\text{orb}}} \int_0^{\Phi} \mathcal{K}^{\text{o}}_{\mathsf{m}}(\Phi') d\Phi' .
\end{equation}
These transformations modify the time-dependent metric and fluid variables $\lbrace k^{\text{d}}_{tr} , \hat{b}^{\text{q}}_t ,  k^{\text{o}}_{tr} \rbrace$ and $\lbrace v^{\text{d}}_r, v^{\text{o}}_r, v^{\text{d}}, \hat{v}^{\text{q}}, v^{\text{o}} \rbrace$, but leave time-independent ones like $b^{\text{q}}_t$ unchanged.

Implementing these changes in the results of Secs.~\ref{subsec:magquad}, \ref{subsec:magdip} and \ref{subsec:magoct}, the Eulerian velocity perturbation $\delta u^{\alpha}$ can be calculated explicitly for the binary neutron star. The contravariant expression $\delta u^{\alpha}$ is related to the covariant components of $\delta u_{\alpha}$ given in Eq.~\eqref{dumag} as follows:

\begin{subequations}
\begin{eqnarray} \label{contravariantr}
\delta u^r &=& f \left[ \delta u_r - e^{-\psi} \left( p_{tr} - p_{rA} \Omega^{AB} \chi^{\text{d}}_B \right) \right] , \\
\delta u^A &=& r^{-2} e^{-\psi} \Omega^{AB} \left[ e^{\psi} \delta u_B - p_{tB} + p_{BC} \Omega^{CD} \chi^{\text{d}}_D \right] - \frac{1}{2} e^{-3\psi} (1-\omega) p_{tt} \Omega^{AB} \chi^{\text{d}}_{B} . \label{contravariantA}
\end{eqnarray}
\end{subequations}
Expressed in terms of the tidal potentials and the variables of this subsection, one finds that

\begin{subequations} \label{contravariant2}
\begin{eqnarray} \label{contravariant2r}
\delta u^r &=& f  \int_0^{\Phi} \left[ \left( \frac{v^{\text{d}}_r}{t \Omega_{\text{orb}}} - e^{-\psi} \frac{k^{\text{d}}_{tr1}}{\Omega_{\text{orb}}}  \right) \sum_{\mathsf{m}} \mathcal{K}^{\text{d}}_{\mathsf{m}}(\Phi') Y^{1 \mathsf{m}} + \left( \frac{v^{\text{o}}_r}{t \Omega_{\text{orb}}} - e^{-\psi} \frac{k^{\text{o}}_{tr1}}{\Omega_{\text{orb}}} \right) \sum_{\mathsf{m}} \mathcal{K}^{\text{o}}_{\mathsf{m}}(\Phi') Y^{3 \mathsf{m}} \right] d\Phi' , \\
\delta u^A &=& r^{-2} \Omega^{AB} \int_0^{\Phi} \left[ \frac{v^{\text{d}}}{t \Omega_{\text{orb}}} \sum_{\mathsf{m}} \mathcal{K}^{\text{d}}_{\mathsf{m}}(\Phi') Y_B^{1 \mathsf{m}} + \left( \frac{\hat{v}^{\text{q}}}{t \Omega_{\text{orb}}} - e^{-\psi} \frac{\hat{b}^{\text{q}}_{t1}}{\Omega_{\text{orb}}} \right) \sum_{\mathsf{m}} \frac{1}{2} \hat{\mathcal{B}}^{\text{q}}_{\mathsf{m}}(\Phi') X_B^{2 \mathsf{m}} + \frac{v^{\text{o}}}{t \Omega_{\text{orb}}} \sum_{\mathsf{m}} \frac{1}{3} \mathcal{K}^{\text{o}}_{\mathsf{m}}(\Phi') Y_B^{3 \mathsf{m}} \right] d\Phi' \nonumber \\ &&- r^{-2} e^{-\psi} b^{\text{q}}_t \Omega^{AB} \mathcal{B}^{\text{q}}_B . \label{contravariant2A}
\end{eqnarray}
\end{subequations}
Equation \eqref{contravariant2} provides a complete description of the currents induced in a neutron star by the gravitomagnetic tidal field of a binary companion to first order in the tidal deformation and the neutron star's spin. The entire radial component $\delta u^r$ is first order in spin, while the angular components $\delta u^A$ consist of a piece $\delta u^A_{\text{tidal}} := - r^{-2} e^{-\psi} b^{\text{q}}_t \Omega^{AB} \mathcal{B}^{\text{q}}_B$, which is zeroth order in spin, and a first-order piece $\delta u^A_{\text{bilinear}}$. The former piece is a tidal current directly induced by the gravitomagnetic quadrupole moment $\mathcal{B}_{ab}$; its time dependence is merely parametric, in the sense that it simply reflects the modulation of $\mathcal{B}_{ab}$ with the orbital phase $\Phi$. The latter piece is a genuinely dynamical tidal current induced by the bilinear spin-coupled gravitomagnetic moments. The integrals of $\mathcal{K}^{\text{d}}_{\mathsf{m}}$, $\hat{\mathcal{B}}^{\text{q}}_{\mathsf{m}}$ and $\mathcal{K}^{\text{o}}_{\mathsf{m}}$ produce a sinusoidal variation of the amplitude that is superposed on the time dependence inherited from Eq.~\eqref{tidmom}.

\begin{table}
\begin{ruledtabular}
\begin{tabular}{dddd}
\multicolumn{1}{c}{\mbox{$n$}} & \multicolumn{1}{c}{\mbox{$M/R$}} & \multicolumn{1}{c}{\mbox{$\sigma$}} & \multicolumn{1}{c}{\mbox{$\delta v~(\text{km/s})$}} \\ \hline
  \noalign{\medskip}
  & 0.13 & 1.117 & 2.234 \\
0.5 & 0.17 & 0.998 & 1.997 \\
  & 0.21 & 0.871 & 1.743 \\
  \noalign{\medskip}
  & 0.13 & 1.233 & 2.466 \\
0.75 & 0.17 & 1.096 & 2.191 \\
  & 0.21 & 0.952 & 1.903 \\
  \noalign{\medskip}
  & 0.13 & 1.327 & 2.654 \\
1.0 & 0.17 & 1.174 & 2.348 \\
  & 0.21 & 1.009 & 2.018
\end{tabular}
\end{ruledtabular}
\caption{\label{table} The dimensionless, equation-of-state dependent parameter $\sigma$ from Eq.~\eqref{sigma} for polytropic models of index $n$ and compactness $M/R$, for use in the estimate Eq.~\eqref{preresult}. The adoption of the fiducial values of Eq.~\eqref{result} for the binary system produces the values of the equatorial velocity perturbation $\delta v$ listed here.}
\end{table}

For concreteness, I now proceed to calculate the amplitude of the tidal currents at the neutron star's equator ($r=R$, $\theta=\pi/2$, $\phi$) by evaluating the tangential velocity perturbation $\delta v^A := R \, \delta u^A$. The radial component of the velocity perturbation automatically vanishes at the surface because the functions $k^{\text{d}}_{tr1}$ and $k^{\text{o}}_{tr1}$, as well as their first derivatives, are zero at $r=R$. The function $\hat{b}^{\text{q}}_{t1}$ also vanishes at $r=R$, and only the $\phi$ component of $\delta v^A$ is nonzero at the equator. Thus, on the basis of Eq.~\eqref{contravariant2}, I compute

\begin{equation} \label{tangentialv0}
\delta v_{\text{tidal}}^{\phi} = - \frac{3 M' R^2 \Omega_{\text{orb}}}{b^2} e^{-\psi} \frac{b^{\text{q}}_t}{R^3} \cos{\left( \phi - \Phi \right)}
\end{equation}
at zeroth order in spin, and

\begin{eqnarray} \label{tangentialv}
\delta v_{\text{bilinear}}^{\phi} = - \frac{M' R^2 \Omega}{b^2} \sigma \sin{\left( \Phi/2 \right)} \sin{\left( \phi - \Phi/2 \right)}
\end{eqnarray}
at first order. I have defined the dimensionless quantity

\begin{equation} \label{sigma}
\sigma := -6  \left(\frac{v^{\text{d}}}{t R^3} + \frac{\hat{v}^{\text{q}}}{t R^3} - \frac{2}{15} \frac{v^{\text{o}}}{t R^3} \right)
\end{equation}
to encode all the equation-of-state dependence of $\delta v^{\phi}_{\text{bilinear}}$. As shown in Table~\ref{table}, $\sigma$ is roughly of order unity for the polytropic models studied in this paper. Restoring factors of $G$ and $c$, the amplitude $\delta v$ of Eq.~\eqref{tangentialv} can be written as

\begin{equation} \label{preresult}
\delta v = \sigma \frac{(2\pi)^{7/3} G^{1/3}}{c^2} \frac{M'}{(M+M')^{2/3}} \frac{R^2}{P} f^{4/3}
\end{equation}
in terms of the neutron star's rotational period $P := 2\pi/\Omega$ and the orbital frequency $f := \Omega_{\text{orb}}/2\pi$.
Evaluated with the parameters of a typical equal-mass neutron star binary system near merger, the dynamical tidal currents have amplitude

\begin{equation} \label{result}
\delta v = 2\sigma \left(\frac{M'}{1.4~M_{\odot}}\right) \left( \frac{2.8~M_{\odot}}{M + M'} \right)^{2/3} \left( \frac{R}{12~\text{km}} \right)^2 \left( \frac{100~\text{ms}}{P} \right) \left( \frac{f}{100~\text{Hz}} \right)^{4/3} \text{km/s}
\end{equation}
at the equator. This figure differs only by a factor of $\sigma$ from the post-Newtonian order-of-magnitude estimate of Poisson and Douc\c ot \cite{Poisson_PNDynResp}.

\begin{acknowledgments}
I am deeply indebted to Eric Poisson for very helpful discussions and advice about this work. I thank J\'er\'emie Gagnon-Bischoff for pointing out a number of typographical errors in an earlier version of the manuscript, and for independently verifying the numerical results displayed in Fig.~\ref{fig:fo}. I also thank John Friedman and Raissa Mendes for useful conversations, as well as Paolo Pani for facilitating the comparison of results with \citex{Pani_TidRotLN}. This work was supported by the Natural Sciences and Engineering Research Council of Canada.
\end{acknowledgments}

\bibliography{Landry-TidDefSlowRotLNS}

\begin{thebibliography}{40}%
\makeatletter
\providecommand \@ifxundefined [1]{%
 \@ifx{#1\undefined}
}%
\providecommand \@ifnum [1]{%
 \ifnum #1\expandafter \@firstoftwo
 \else \expandafter \@secondoftwo
 \fi
}%
\providecommand \@ifx [1]{%
 \ifx #1\expandafter \@firstoftwo
 \else \expandafter \@secondoftwo
 \fi
}%
\providecommand \natexlab [1]{#1}%
\providecommand \enquote  [1]{``#1''}%
\providecommand \bibnamefont  [1]{#1}%
\providecommand \bibfnamefont [1]{#1}%
\providecommand \citenamefont [1]{#1}%
\providecommand \href@noop [0]{\@secondoftwo}%
\providecommand \href [0]{\begingroup \@sanitize@url \@href}%
\providecommand \@href[1]{\@@startlink{#1}\@@href}%
\providecommand \@@href[1]{\endgroup#1\@@endlink}%
\providecommand \@sanitize@url [0]{\catcode `\\12\catcode `\$12\catcode
  `\&12\catcode `\#12\catcode `\^12\catcode `\_12\catcode `\%12\relax}%
\providecommand \@@startlink[1]{}%
\providecommand \@@endlink[0]{}%
\providecommand \url  [0]{\begingroup\@sanitize@url \@url }%
\providecommand \@url [1]{\endgroup\@href {#1}{\urlprefix }}%
\providecommand \urlprefix  [0]{URL }%
\providecommand \Eprint [0]{\href }%
\providecommand \doibase [0]{http://dx.doi.org/}%
\providecommand \selectlanguage [0]{\@gobble}%
\providecommand \bibinfo  [0]{\@secondoftwo}%
\providecommand \bibfield  [0]{\@secondoftwo}%
\providecommand \translation [1]{[#1]}%
\providecommand \BibitemOpen [0]{}%
\providecommand \bibitemStop [0]{}%
\providecommand \bibitemNoStop [0]{.\EOS\space}%
\providecommand \EOS [0]{\spacefactor3000\relax}%
\providecommand \BibitemShut  [1]{\csname bibitem#1\endcsname}%
\let\auto@bib@innerbib\@empty
\bibitem [{\citenamefont {{Poisson}}(2015)}]{Poisson_TidDefRotBH}%
  \BibitemOpen
  \bibfield  {author} {\bibinfo {author} {\bibfnamefont {E.}~\bibnamefont
  {{Poisson}}},\ }\href {\doibase 10.1103/PhysRevD.91.044004} {\bibfield
  {journal} {\bibinfo  {journal} {\prd}\ }\textbf {\bibinfo {volume} {91}},\
  \bibinfo {eid} {044004} (\bibinfo {year} {2015})},\ \Eprint
  {http://arxiv.org/abs/1411.4711} {arXiv:1411.4711 [gr-qc]} \BibitemShut
  {NoStop}%
\bibitem [{\citenamefont {{Landry}}\ and\ \citenamefont
  {{Poisson}}(2015{\natexlab{a}})}]{Landry_TidDefRotExt}%
  \BibitemOpen
  \bibfield  {author} {\bibinfo {author} {\bibfnamefont {P.}~\bibnamefont
  {{Landry}}}\ and\ \bibinfo {author} {\bibfnamefont {E.}~\bibnamefont
  {{Poisson}}},\ }\href {\doibase 10.1103/PhysRevD.91.104018} {\bibfield
  {journal} {\bibinfo  {journal} {\prd}\ }\textbf {\bibinfo {volume} {91}},\
  \bibinfo {eid} {104018} (\bibinfo {year} {2015}{\natexlab{a}})},\ \Eprint
  {http://arxiv.org/abs/1503.07366} {arXiv:1503.07366 [gr-qc]} \BibitemShut
  {NoStop}%
\bibitem [{\citenamefont {{Pani}}\ \emph
  {et~al.}(2015{\natexlab{a}})\citenamefont {{Pani}}, \citenamefont
  {{Gualtieri}}, \citenamefont {{Maselli}},\ and\ \citenamefont
  {{Ferrari}}}]{Pani_TidDefSpin}%
  \BibitemOpen
  \bibfield  {author} {\bibinfo {author} {\bibfnamefont {P.}~\bibnamefont
  {{Pani}}}, \bibinfo {author} {\bibfnamefont {L.}~\bibnamefont {{Gualtieri}}},
  \bibinfo {author} {\bibfnamefont {A.}~\bibnamefont {{Maselli}}}, \ and\
  \bibinfo {author} {\bibfnamefont {V.}~\bibnamefont {{Ferrari}}},\ }\href
  {\doibase 10.1103/PhysRevD.92.024010} {\bibfield  {journal} {\bibinfo
  {journal} {\prd}\ }\textbf {\bibinfo {volume} {92}},\ \bibinfo {eid} {024010}
  (\bibinfo {year} {2015}{\natexlab{a}})},\ \Eprint
  {http://arxiv.org/abs/1503.07365} {arXiv:1503.07365 [gr-qc]} \BibitemShut
  {NoStop}%
\bibitem [{\citenamefont {{Pani}}\ \emph
  {et~al.}(2015{\natexlab{b}})\citenamefont {{Pani}}, \citenamefont
  {{Gualtieri}},\ and\ \citenamefont {{Ferrari}}}]{Pani_TidRotLN}%
  \BibitemOpen
  \bibfield  {author} {\bibinfo {author} {\bibfnamefont {P.}~\bibnamefont
  {{Pani}}}, \bibinfo {author} {\bibfnamefont {L.}~\bibnamefont {{Gualtieri}}},
  \ and\ \bibinfo {author} {\bibfnamefont {V.}~\bibnamefont {{Ferrari}}},\
  }\href {\doibase 10.1103/PhysRevD.92.124003} {\bibfield  {journal} {\bibinfo
  {journal} {\prd}\ }\textbf {\bibinfo {volume} {92}},\ \bibinfo {eid} {124003}
  (\bibinfo {year} {2015}{\natexlab{b}})},\ \Eprint
  {http://arxiv.org/abs/1509.02171} {arXiv:1509.02171 [gr-qc]} \BibitemShut
  {NoStop}%
\bibitem [{\citenamefont {{Landry}}\ and\ \citenamefont
  {{Poisson}}(2015{\natexlab{b}})}]{Landry_DynResp}%
  \BibitemOpen
  \bibfield  {author} {\bibinfo {author} {\bibfnamefont {P.}~\bibnamefont
  {{Landry}}}\ and\ \bibinfo {author} {\bibfnamefont {E.}~\bibnamefont
  {{Poisson}}},\ }\href {\doibase 10.1103/PhysRevD.92.124041} {\bibfield
  {journal} {\bibinfo  {journal} {\prd}\ }\textbf {\bibinfo {volume} {92}},\
  \bibinfo {eid} {124041} (\bibinfo {year} {2015}{\natexlab{b}})},\ \Eprint
  {http://arxiv.org/abs/1510.09170} {arXiv:1510.09170 [gr-qc]} \BibitemShut
  {NoStop}%
\bibitem [{\citenamefont {Poisson}\ and\ \citenamefont
  {Dou\ifmmode~\mbox{\c{c}}\else \c{c}\fi{}ot}(2017)}]{Poisson_PNDynResp}%
  \BibitemOpen
  \bibfield  {author} {\bibinfo {author} {\bibfnamefont {E.}~\bibnamefont
  {Poisson}}\ and\ \bibinfo {author} {\bibfnamefont {J.}~\bibnamefont
  {Dou\ifmmode~\mbox{\c{c}}\else \c{c}\fi{}ot}},\ }\href {\doibase
  10.1103/PhysRevD.95.044023} {\bibfield  {journal} {\bibinfo  {journal} {Phys.
  Rev. D}\ }\textbf {\bibinfo {volume} {95}},\ \bibinfo {pages} {044023}
  (\bibinfo {year} {2017})},\ \Eprint {http://arxiv.org/abs/1612.04255}
  {arXiv:1612.04255 [gr-qc]} \BibitemShut {NoStop}%
\bibitem [{\citenamefont {{Flanagan}}\ and\ \citenamefont
  {{Hinderer}}(2008)}]{Flanagan_NSLNGW}%
  \BibitemOpen
  \bibfield  {author} {\bibinfo {author} {\bibfnamefont {{\'E}.~{\'E}.}\
  \bibnamefont {{Flanagan}}}\ and\ \bibinfo {author} {\bibfnamefont
  {T.}~\bibnamefont {{Hinderer}}},\ }\href {\doibase
  10.1103/PhysRevD.77.021502} {\bibfield  {journal} {\bibinfo  {journal}
  {\prd}\ }\textbf {\bibinfo {volume} {77}},\ \bibinfo {eid} {021502} (\bibinfo
  {year} {2008})},\ \Eprint {http://arxiv.org/abs/0709.1915} {arXiv:0709.1915}
  \BibitemShut {NoStop}%
\bibitem [{\citenamefont {{Hinderer}}(2008)}]{Hinderer_NSLN}%
  \BibitemOpen
  \bibfield  {author} {\bibinfo {author} {\bibfnamefont {T.}~\bibnamefont
  {{Hinderer}}},\ }\href {\doibase 10.1086/533487} {\bibfield  {journal}
  {\bibinfo  {journal} {\apj}\ }\textbf {\bibinfo {volume} {677}},\ \bibinfo
  {eid} {1216-1220} (\bibinfo {year} {2008})},\ \Eprint
  {http://arxiv.org/abs/0711.2420} {arXiv:0711.2420} \BibitemShut {NoStop}%
\bibitem [{\citenamefont {{Damour}}\ and\ \citenamefont
  {{Nagar}}(2009)}]{Damour_TidPropNS}%
  \BibitemOpen
  \bibfield  {author} {\bibinfo {author} {\bibfnamefont {T.}~\bibnamefont
  {{Damour}}}\ and\ \bibinfo {author} {\bibfnamefont {A.}~\bibnamefont
  {{Nagar}}},\ }\href {\doibase 10.1103/PhysRevD.80.084035} {\bibfield
  {journal} {\bibinfo  {journal} {\prd}\ }\textbf {\bibinfo {volume} {80}},\
  \bibinfo {eid} {084035} (\bibinfo {year} {2009})},\ \Eprint
  {http://arxiv.org/abs/0906.0096} {arXiv:0906.0096 [gr-qc]} \BibitemShut
  {NoStop}%
\bibitem [{\citenamefont {{Binnington}}\ and\ \citenamefont
  {{Poisson}}(2009)}]{Binnington}%
  \BibitemOpen
  \bibfield  {author} {\bibinfo {author} {\bibfnamefont {T.}~\bibnamefont
  {{Binnington}}}\ and\ \bibinfo {author} {\bibfnamefont {E.}~\bibnamefont
  {{Poisson}}},\ }\href {\doibase 10.1103/PhysRevD.80.084018} {\bibfield
  {journal} {\bibinfo  {journal} {\prd}\ }\textbf {\bibinfo {volume} {80}},\
  \bibinfo {eid} {084018} (\bibinfo {year} {2009})},\ \Eprint
  {http://arxiv.org/abs/0906.1366} {arXiv:0906.1366 [gr-qc]} \BibitemShut
  {NoStop}%
\bibitem [{\citenamefont {{Landry}}\ and\ \citenamefont
  {{Poisson}}(2014)}]{Landry_SurfLN}%
  \BibitemOpen
  \bibfield  {author} {\bibinfo {author} {\bibfnamefont {P.}~\bibnamefont
  {{Landry}}}\ and\ \bibinfo {author} {\bibfnamefont {E.}~\bibnamefont
  {{Poisson}}},\ }\href {\doibase 10.1103/PhysRevD.89.124011} {\bibfield
  {journal} {\bibinfo  {journal} {\prd}\ }\textbf {\bibinfo {volume} {89}},\
  \bibinfo {eid} {124011} (\bibinfo {year} {2014})},\ \Eprint
  {http://arxiv.org/abs/1404.6798} {arXiv:1404.6798 [gr-qc]} \BibitemShut
  {NoStop}%
\bibitem [{\citenamefont {{Hinderer}}\ \emph {et~al.}(2010)\citenamefont
  {{Hinderer}}, \citenamefont {{Lackey}}, \citenamefont {{Lang}},\ and\
  \citenamefont {{Read}}}]{Hinderer_TidDefNS}%
  \BibitemOpen
  \bibfield  {author} {\bibinfo {author} {\bibfnamefont {T.}~\bibnamefont
  {{Hinderer}}}, \bibinfo {author} {\bibfnamefont {B.~D.}\ \bibnamefont
  {{Lackey}}}, \bibinfo {author} {\bibfnamefont {R.~N.}\ \bibnamefont
  {{Lang}}}, \ and\ \bibinfo {author} {\bibfnamefont {J.~S.}\ \bibnamefont
  {{Read}}},\ }\href {\doibase 10.1103/PhysRevD.81.123016} {\bibfield
  {journal} {\bibinfo  {journal} {\prd}\ }\textbf {\bibinfo {volume} {81}},\
  \bibinfo {eid} {123016} (\bibinfo {year} {2010})},\ \Eprint
  {http://arxiv.org/abs/0911.3535} {arXiv:0911.3535 [astro-ph.HE]} \BibitemShut
  {NoStop}%
\bibitem [{\citenamefont {{Baiotti}}\ \emph {et~al.}(2010)\citenamefont
  {{Baiotti}}, \citenamefont {{Damour}}, \citenamefont {{Giacomazzo}},
  \citenamefont {{Nagar}},\ and\ \citenamefont
  {{Rezzolla}}}]{Baiotti_TidEffInsp}%
  \BibitemOpen
  \bibfield  {author} {\bibinfo {author} {\bibfnamefont {L.}~\bibnamefont
  {{Baiotti}}}, \bibinfo {author} {\bibfnamefont {T.}~\bibnamefont {{Damour}}},
  \bibinfo {author} {\bibfnamefont {B.}~\bibnamefont {{Giacomazzo}}}, \bibinfo
  {author} {\bibfnamefont {A.}~\bibnamefont {{Nagar}}}, \ and\ \bibinfo
  {author} {\bibfnamefont {L.}~\bibnamefont {{Rezzolla}}},\ }\href {\doibase
  10.1103/PhysRevLett.105.261101} {\bibfield  {journal} {\bibinfo  {journal}
  {Physical Review Letters}\ }\textbf {\bibinfo {volume} {105}},\ \bibinfo
  {eid} {261101} (\bibinfo {year} {2010})},\ \Eprint
  {http://arxiv.org/abs/1009.0521} {arXiv:1009.0521 [gr-qc]} \BibitemShut
  {NoStop}%
\bibitem [{\citenamefont {{Baiotti}}\ \emph {et~al.}(2011)\citenamefont
  {{Baiotti}}, \citenamefont {{Damour}}, \citenamefont {{Giacomazzo}},
  \citenamefont {{Nagar}},\ and\ \citenamefont
  {{Rezzolla}}}]{Baiotti_InspBinNS}%
  \BibitemOpen
  \bibfield  {author} {\bibinfo {author} {\bibfnamefont {L.}~\bibnamefont
  {{Baiotti}}}, \bibinfo {author} {\bibfnamefont {T.}~\bibnamefont {{Damour}}},
  \bibinfo {author} {\bibfnamefont {B.}~\bibnamefont {{Giacomazzo}}}, \bibinfo
  {author} {\bibfnamefont {A.}~\bibnamefont {{Nagar}}}, \ and\ \bibinfo
  {author} {\bibfnamefont {L.}~\bibnamefont {{Rezzolla}}},\ }\href {\doibase
  10.1103/PhysRevD.84.024017} {\bibfield  {journal} {\bibinfo  {journal}
  {\prd}\ }\textbf {\bibinfo {volume} {84}},\ \bibinfo {eid} {024017} (\bibinfo
  {year} {2011})},\ \Eprint {http://arxiv.org/abs/1103.3874} {arXiv:1103.3874
  [gr-qc]} \BibitemShut {NoStop}%
\bibitem [{\citenamefont {{Vines}}\ \emph {et~al.}(2011)\citenamefont
  {{Vines}}, \citenamefont {{Flanagan}},\ and\ \citenamefont
  {{Hinderer}}}]{Vines_PNTidEff}%
  \BibitemOpen
  \bibfield  {author} {\bibinfo {author} {\bibfnamefont {J.}~\bibnamefont
  {{Vines}}}, \bibinfo {author} {\bibfnamefont {{\'E}.~{\'E}.}\ \bibnamefont
  {{Flanagan}}}, \ and\ \bibinfo {author} {\bibfnamefont {T.}~\bibnamefont
  {{Hinderer}}},\ }\href {\doibase 10.1103/PhysRevD.83.084051} {\bibfield
  {journal} {\bibinfo  {journal} {\prd}\ }\textbf {\bibinfo {volume} {83}},\
  \bibinfo {eid} {084051} (\bibinfo {year} {2011})},\ \Eprint
  {http://arxiv.org/abs/1101.1673} {arXiv:1101.1673 [gr-qc]} \BibitemShut
  {NoStop}%
\bibitem [{\citenamefont {{Pannarale}}\ \emph {et~al.}(2011)\citenamefont
  {{Pannarale}}, \citenamefont {{Rezzolla}}, \citenamefont {{Ohme}},\ and\
  \citenamefont {{Read}}}]{Pannarale}%
  \BibitemOpen
  \bibfield  {author} {\bibinfo {author} {\bibfnamefont {F.}~\bibnamefont
  {{Pannarale}}}, \bibinfo {author} {\bibfnamefont {L.}~\bibnamefont
  {{Rezzolla}}}, \bibinfo {author} {\bibfnamefont {F.}~\bibnamefont {{Ohme}}},
  \ and\ \bibinfo {author} {\bibfnamefont {J.~S.}\ \bibnamefont {{Read}}},\
  }\href {\doibase 10.1103/PhysRevD.84.104017} {\bibfield  {journal} {\bibinfo
  {journal} {\prd}\ }\textbf {\bibinfo {volume} {84}},\ \bibinfo {eid} {104017}
  (\bibinfo {year} {2011})},\ \Eprint {http://arxiv.org/abs/1103.3526}
  {arXiv:1103.3526 [astro-ph.HE]} \BibitemShut {NoStop}%
\bibitem [{\citenamefont {{Lackey}}\ \emph {et~al.}(2012)\citenamefont
  {{Lackey}}, \citenamefont {{Kyutoku}}, \citenamefont {{Shibata}},
  \citenamefont {{Brady}},\ and\ \citenamefont
  {{Friedman}}}]{Lackey_NonSpinBH}%
  \BibitemOpen
  \bibfield  {author} {\bibinfo {author} {\bibfnamefont {B.~D.}\ \bibnamefont
  {{Lackey}}}, \bibinfo {author} {\bibfnamefont {K.}~\bibnamefont {{Kyutoku}}},
  \bibinfo {author} {\bibfnamefont {M.}~\bibnamefont {{Shibata}}}, \bibinfo
  {author} {\bibfnamefont {P.~R.}\ \bibnamefont {{Brady}}}, \ and\ \bibinfo
  {author} {\bibfnamefont {J.~L.}\ \bibnamefont {{Friedman}}},\ }\href
  {\doibase 10.1103/PhysRevD.85.044061} {\bibfield  {journal} {\bibinfo
  {journal} {\prd}\ }\textbf {\bibinfo {volume} {85}},\ \bibinfo {eid} {044061}
  (\bibinfo {year} {2012})},\ \Eprint {http://arxiv.org/abs/1109.3402}
  {arXiv:1109.3402 [astro-ph.HE]} \BibitemShut {NoStop}%
\bibitem [{\citenamefont {{Damour}}\ \emph {et~al.}(2012)\citenamefont
  {{Damour}}, \citenamefont {{Nagar}},\ and\ \citenamefont
  {{Villain}}}]{Damour_TidPolarNS}%
  \BibitemOpen
  \bibfield  {author} {\bibinfo {author} {\bibfnamefont {T.}~\bibnamefont
  {{Damour}}}, \bibinfo {author} {\bibfnamefont {A.}~\bibnamefont {{Nagar}}}, \
  and\ \bibinfo {author} {\bibfnamefont {L.}~\bibnamefont {{Villain}}},\ }\href
  {\doibase 10.1103/PhysRevD.85.123007} {\bibfield  {journal} {\bibinfo
  {journal} {\prd}\ }\textbf {\bibinfo {volume} {85}},\ \bibinfo {eid} {123007}
  (\bibinfo {year} {2012})},\ \Eprint {http://arxiv.org/abs/1203.4352}
  {arXiv:1203.4352 [gr-qc]} \BibitemShut {NoStop}%
\bibitem [{\citenamefont {{Read}}\ \emph {et~al.}(2013)\citenamefont {{Read}},
  \citenamefont {{Baiotti}}, \citenamefont {{Creighton}}, \citenamefont
  {{Friedman}}, \citenamefont {{Giacomazzo}}, \citenamefont {{Kyutoku}},
  \citenamefont {{Markakis}}, \citenamefont {{Rezzolla}}, \citenamefont
  {{Shibata}},\ and\ \citenamefont {{Taniguchi}}}]{Read_MatterEffGW}%
  \BibitemOpen
  \bibfield  {author} {\bibinfo {author} {\bibfnamefont {J.~S.}\ \bibnamefont
  {{Read}}}, \bibinfo {author} {\bibfnamefont {L.}~\bibnamefont {{Baiotti}}},
  \bibinfo {author} {\bibfnamefont {J.~D.~E.}\ \bibnamefont {{Creighton}}},
  \bibinfo {author} {\bibfnamefont {J.~L.}\ \bibnamefont {{Friedman}}},
  \bibinfo {author} {\bibfnamefont {B.}~\bibnamefont {{Giacomazzo}}}, \bibinfo
  {author} {\bibfnamefont {K.}~\bibnamefont {{Kyutoku}}}, \bibinfo {author}
  {\bibfnamefont {C.}~\bibnamefont {{Markakis}}}, \bibinfo {author}
  {\bibfnamefont {L.}~\bibnamefont {{Rezzolla}}}, \bibinfo {author}
  {\bibfnamefont {M.}~\bibnamefont {{Shibata}}}, \ and\ \bibinfo {author}
  {\bibfnamefont {K.}~\bibnamefont {{Taniguchi}}},\ }\href {\doibase
  10.1103/PhysRevD.88.044042} {\bibfield  {journal} {\bibinfo  {journal}
  {\prd}\ }\textbf {\bibinfo {volume} {88}},\ \bibinfo {eid} {044042} (\bibinfo
  {year} {2013})},\ \Eprint {http://arxiv.org/abs/1306.4065} {arXiv:1306.4065
  [gr-qc]} \BibitemShut {NoStop}%
\bibitem [{\citenamefont {{Vines}}\ and\ \citenamefont
  {{Flanagan}}(2013)}]{Vines_PNQuadTid}%
  \BibitemOpen
  \bibfield  {author} {\bibinfo {author} {\bibfnamefont {J.~E.}\ \bibnamefont
  {{Vines}}}\ and\ \bibinfo {author} {\bibfnamefont {{\'E}.~{\'E}.}\
  \bibnamefont {{Flanagan}}},\ }\href {\doibase 10.1103/PhysRevD.88.024046}
  {\bibfield  {journal} {\bibinfo  {journal} {\prd}\ }\textbf {\bibinfo
  {volume} {88}},\ \bibinfo {eid} {024046} (\bibinfo {year} {2013})},\ \Eprint
  {http://arxiv.org/abs/1009.4919} {arXiv:1009.4919 [gr-qc]} \BibitemShut
  {NoStop}%
\bibitem [{\citenamefont {{Maselli}}\ \emph
  {et~al.}(2013{\natexlab{a}})\citenamefont {{Maselli}}, \citenamefont
  {{Gualtieri}},\ and\ \citenamefont {{Ferrari}}}]{Maselli_NSEoSGW}%
  \BibitemOpen
  \bibfield  {author} {\bibinfo {author} {\bibfnamefont {A.}~\bibnamefont
  {{Maselli}}}, \bibinfo {author} {\bibfnamefont {L.}~\bibnamefont
  {{Gualtieri}}}, \ and\ \bibinfo {author} {\bibfnamefont {V.}~\bibnamefont
  {{Ferrari}}},\ }\href {\doibase 10.1103/PhysRevD.88.104040} {\bibfield
  {journal} {\bibinfo  {journal} {\prd}\ }\textbf {\bibinfo {volume} {88}},\
  \bibinfo {eid} {104040} (\bibinfo {year} {2013}{\natexlab{a}})},\ \Eprint
  {http://arxiv.org/abs/1310.5381} {arXiv:1310.5381 [gr-qc]} \BibitemShut
  {NoStop}%
\bibitem [{\citenamefont {{Lackey}}\ \emph {et~al.}(2014)\citenamefont
  {{Lackey}}, \citenamefont {{Kyutoku}}, \citenamefont {{Shibata}},
  \citenamefont {{Brady}},\ and\ \citenamefont
  {{Friedman}}}]{Lackey_AlignedSpinBH}%
  \BibitemOpen
  \bibfield  {author} {\bibinfo {author} {\bibfnamefont {B.~D.}\ \bibnamefont
  {{Lackey}}}, \bibinfo {author} {\bibfnamefont {K.}~\bibnamefont {{Kyutoku}}},
  \bibinfo {author} {\bibfnamefont {M.}~\bibnamefont {{Shibata}}}, \bibinfo
  {author} {\bibfnamefont {P.~R.}\ \bibnamefont {{Brady}}}, \ and\ \bibinfo
  {author} {\bibfnamefont {J.~L.}\ \bibnamefont {{Friedman}}},\ }\href
  {\doibase 10.1103/PhysRevD.89.043009} {\bibfield  {journal} {\bibinfo
  {journal} {\prd}\ }\textbf {\bibinfo {volume} {89}},\ \bibinfo {eid} {043009}
  (\bibinfo {year} {2014})},\ \Eprint {http://arxiv.org/abs/1303.6298}
  {arXiv:1303.6298 [gr-qc]} \BibitemShut {NoStop}%
\bibitem [{\citenamefont {{Favata}}(2014)}]{Favata_ParamErr}%
  \BibitemOpen
  \bibfield  {author} {\bibinfo {author} {\bibfnamefont {M.}~\bibnamefont
  {{Favata}}},\ }\href {\doibase 10.1103/PhysRevLett.112.101101} {\bibfield
  {journal} {\bibinfo  {journal} {Physical Review Letters}\ }\textbf {\bibinfo
  {volume} {112}},\ \bibinfo {eid} {101101} (\bibinfo {year} {2014})},\ \Eprint
  {http://arxiv.org/abs/1310.8288} {arXiv:1310.8288 [gr-qc]} \BibitemShut
  {NoStop}%
\bibitem [{\citenamefont {{Yagi}}\ and\ \citenamefont
  {{Yunes}}(2014)}]{Yagi_HardToMeasure}%
  \BibitemOpen
  \bibfield  {author} {\bibinfo {author} {\bibfnamefont {K.}~\bibnamefont
  {{Yagi}}}\ and\ \bibinfo {author} {\bibfnamefont {N.}~\bibnamefont
  {{Yunes}}},\ }\href {\doibase 10.1103/PhysRevD.89.021303} {\bibfield
  {journal} {\bibinfo  {journal} {\prd}\ }\textbf {\bibinfo {volume} {89}},\
  \bibinfo {eid} {021303} (\bibinfo {year} {2014})},\ \Eprint
  {http://arxiv.org/abs/1310.8358} {arXiv:1310.8358 [gr-qc]} \BibitemShut
  {NoStop}%
\bibitem [{\citenamefont {{Yagi}}\ and\ \citenamefont
  {{Yunes}}(2013{\natexlab{a}})}]{Yagi_ILoveQ}%
  \BibitemOpen
  \bibfield  {author} {\bibinfo {author} {\bibfnamefont {K.}~\bibnamefont
  {{Yagi}}}\ and\ \bibinfo {author} {\bibfnamefont {N.}~\bibnamefont
  {{Yunes}}},\ }\href {\doibase 10.1126/science.1236462} {\bibfield  {journal}
  {\bibinfo  {journal} {Science}\ }\textbf {\bibinfo {volume} {341}},\ \bibinfo
  {pages} {365} (\bibinfo {year} {2013}{\natexlab{a}})},\ \Eprint
  {http://arxiv.org/abs/1302.4499} {arXiv:1302.4499 [gr-qc]} \BibitemShut
  {NoStop}%
\bibitem [{\citenamefont {{Yagi}}\ and\ \citenamefont
  {{Yunes}}(2013{\natexlab{b}})}]{Yagi_ILoveQApp}%
  \BibitemOpen
  \bibfield  {author} {\bibinfo {author} {\bibfnamefont {K.}~\bibnamefont
  {{Yagi}}}\ and\ \bibinfo {author} {\bibfnamefont {N.}~\bibnamefont
  {{Yunes}}},\ }\href {\doibase 10.1103/PhysRevD.88.023009} {\bibfield
  {journal} {\bibinfo  {journal} {\prd}\ }\textbf {\bibinfo {volume} {88}},\
  \bibinfo {eid} {023009} (\bibinfo {year} {2013}{\natexlab{b}})},\ \Eprint
  {http://arxiv.org/abs/1303.1528} {arXiv:1303.1528 [gr-qc]} \BibitemShut
  {NoStop}%
\bibitem [{\citenamefont {{Delsate}}(2015)}]{Delsate_ILove}%
  \BibitemOpen
  \bibfield  {author} {\bibinfo {author} {\bibfnamefont {T.}~\bibnamefont
  {{Delsate}}},\ }\href {\doibase 10.1103/PhysRevD.92.124001} {\bibfield
  {journal} {\bibinfo  {journal} {\prd}\ }\textbf {\bibinfo {volume} {92}},\
  \bibinfo {eid} {124001} (\bibinfo {year} {2015})},\ \Eprint
  {http://arxiv.org/abs/1504.07335} {arXiv:1504.07335 [gr-qc]} \BibitemShut
  {NoStop}%
\bibitem [{\citenamefont {{Doneva}}\ \emph {et~al.}(2014)\citenamefont
  {{Doneva}}, \citenamefont {{Yazadjiev}}, \citenamefont {{Stergioulas}},\ and\
  \citenamefont {{Kokkotas}}}]{Doneva}%
  \BibitemOpen
  \bibfield  {author} {\bibinfo {author} {\bibfnamefont {D.~D.}\ \bibnamefont
  {{Doneva}}}, \bibinfo {author} {\bibfnamefont {S.~S.}\ \bibnamefont
  {{Yazadjiev}}}, \bibinfo {author} {\bibfnamefont {N.}~\bibnamefont
  {{Stergioulas}}}, \ and\ \bibinfo {author} {\bibfnamefont {K.~D.}\
  \bibnamefont {{Kokkotas}}},\ }\href {\doibase 10.1088/2041-8205/781/1/L6}
  {\bibfield  {journal} {\bibinfo  {journal} {\apjl}\ }\textbf {\bibinfo
  {volume} {781}},\ \bibinfo {eid} {L6} (\bibinfo {year} {2014})},\ \Eprint
  {http://arxiv.org/abs/1310.7436} {arXiv:1310.7436 [gr-qc]} \BibitemShut
  {NoStop}%
\bibitem [{\citenamefont {{Maselli}}\ \emph
  {et~al.}(2013{\natexlab{b}})\citenamefont {{Maselli}}, \citenamefont
  {{Cardoso}}, \citenamefont {{Ferrari}}, \citenamefont {{Gualtieri}},\ and\
  \citenamefont {{Pani}}}]{Maselli_EoSIndepRelNS}%
  \BibitemOpen
  \bibfield  {author} {\bibinfo {author} {\bibfnamefont {A.}~\bibnamefont
  {{Maselli}}}, \bibinfo {author} {\bibfnamefont {V.}~\bibnamefont
  {{Cardoso}}}, \bibinfo {author} {\bibfnamefont {V.}~\bibnamefont
  {{Ferrari}}}, \bibinfo {author} {\bibfnamefont {L.}~\bibnamefont
  {{Gualtieri}}}, \ and\ \bibinfo {author} {\bibfnamefont {P.}~\bibnamefont
  {{Pani}}},\ }\href {\doibase 10.1103/PhysRevD.88.023007} {\bibfield
  {journal} {\bibinfo  {journal} {\prd}\ }\textbf {\bibinfo {volume} {88}},\
  \bibinfo {eid} {023007} (\bibinfo {year} {2013}{\natexlab{b}})},\ \Eprint
  {http://arxiv.org/abs/1304.2052} {arXiv:1304.2052 [gr-qc]} \BibitemShut
  {NoStop}%
\bibitem [{\citenamefont {{Yagi}}(2014)}]{Yagi_MultiLove}%
  \BibitemOpen
  \bibfield  {author} {\bibinfo {author} {\bibfnamefont {K.}~\bibnamefont
  {{Yagi}}},\ }\href {\doibase 10.1103/PhysRevD.89.043011} {\bibfield
  {journal} {\bibinfo  {journal} {\prd}\ }\textbf {\bibinfo {volume} {89}},\
  \bibinfo {eid} {043011} (\bibinfo {year} {2014})},\ \Eprint
  {http://arxiv.org/abs/1311.0872} {arXiv:1311.0872 [gr-qc]} \BibitemShut
  {NoStop}%
\bibitem [{\citenamefont {{Haskell}}\ \emph {et~al.}(2014)\citenamefont
  {{Haskell}}, \citenamefont {{Ciolfi}}, \citenamefont {{Pannarale}},\ and\
  \citenamefont {{Rezzolla}}}]{Haskell}%
  \BibitemOpen
  \bibfield  {author} {\bibinfo {author} {\bibfnamefont {B.}~\bibnamefont
  {{Haskell}}}, \bibinfo {author} {\bibfnamefont {R.}~\bibnamefont {{Ciolfi}}},
  \bibinfo {author} {\bibfnamefont {F.}~\bibnamefont {{Pannarale}}}, \ and\
  \bibinfo {author} {\bibfnamefont {L.}~\bibnamefont {{Rezzolla}}},\ }\href
  {\doibase 10.1093/mnrasl/slt161} {\bibfield  {journal} {\bibinfo  {journal}
  {\mnras}\ }\textbf {\bibinfo {volume} {438}},\ \bibinfo {pages} {L71}
  (\bibinfo {year} {2014})},\ \Eprint {http://arxiv.org/abs/1309.3885}
  {arXiv:1309.3885 [astro-ph.SR]} \BibitemShut {NoStop}%
\bibitem [{\citenamefont {{Chakrabarti}}\ \emph {et~al.}(2014)\citenamefont
  {{Chakrabarti}}, \citenamefont {{Delsate}}, \citenamefont {{G{\"u}rlebeck}},\
  and\ \citenamefont {{Steinhoff}}}]{Chakrabarti_IQRapidRotNS}%
  \BibitemOpen
  \bibfield  {author} {\bibinfo {author} {\bibfnamefont {S.}~\bibnamefont
  {{Chakrabarti}}}, \bibinfo {author} {\bibfnamefont {T.}~\bibnamefont
  {{Delsate}}}, \bibinfo {author} {\bibfnamefont {N.}~\bibnamefont
  {{G{\"u}rlebeck}}}, \ and\ \bibinfo {author} {\bibfnamefont {J.}~\bibnamefont
  {{Steinhoff}}},\ }\href {\doibase 10.1103/PhysRevLett.112.201102} {\bibfield
  {journal} {\bibinfo  {journal} {Physical Review Letters}\ }\textbf {\bibinfo
  {volume} {112}},\ \bibinfo {eid} {201102} (\bibinfo {year} {2014})},\ \Eprint
  {http://arxiv.org/abs/1311.6509} {arXiv:1311.6509 [gr-qc]} \BibitemShut
  {NoStop}%
\bibitem [{\citenamefont {{Landry}}\ and\ \citenamefont
  {{Poisson}}(2015{\natexlab{c}})}]{Landry_Irrot}%
  \BibitemOpen
  \bibfield  {author} {\bibinfo {author} {\bibfnamefont {P.}~\bibnamefont
  {{Landry}}}\ and\ \bibinfo {author} {\bibfnamefont {E.}~\bibnamefont
  {{Poisson}}},\ }\href {\doibase 10.1103/PhysRevD.91.104026} {\bibfield
  {journal} {\bibinfo  {journal} {\prd}\ }\textbf {\bibinfo {volume} {91}},\
  \bibinfo {eid} {104026} (\bibinfo {year} {2015}{\natexlab{c}})},\ \Eprint
  {http://arxiv.org/abs/1504.06606} {arXiv:1504.06606 [gr-qc]} \BibitemShut
  {NoStop}%
\bibitem [{\citenamefont {{Shapiro}}(1996)}]{Shapiro}%
  \BibitemOpen
  \bibfield  {author} {\bibinfo {author} {\bibfnamefont {S.~L.}\ \bibnamefont
  {{Shapiro}}},\ }\href {\doibase 10.1103/PhysRevLett.77.4487} {\bibfield
  {journal} {\bibinfo  {journal} {Physical Review Letters}\ }\textbf {\bibinfo
  {volume} {77}},\ \bibinfo {pages} {4487} (\bibinfo {year}
  {1996})}\BibitemShut {NoStop}%
\bibitem [{\citenamefont {{Favata}}(2006)}]{Favata_CrushedNS}%
  \BibitemOpen
  \bibfield  {author} {\bibinfo {author} {\bibfnamefont {M.}~\bibnamefont
  {{Favata}}},\ }\href {\doibase 10.1103/PhysRevD.73.104005} {\bibfield
  {journal} {\bibinfo  {journal} {\prd}\ }\textbf {\bibinfo {volume} {73}},\
  \bibinfo {eid} {104005} (\bibinfo {year} {2006})},\ \Eprint
  {http://arxiv.org/abs/astro-ph/0510668} {astro-ph/0510668} \BibitemShut
  {NoStop}%
\bibitem [{\citenamefont {{Tooper}}(1965)}]{Tooper_AdFluidSph}%
  \BibitemOpen
  \bibfield  {author} {\bibinfo {author} {\bibfnamefont {R.~F.}\ \bibnamefont
  {{Tooper}}},\ }\href {\doibase 10.1086/148435} {\bibfield  {journal}
  {\bibinfo  {journal} {\apj}\ }\textbf {\bibinfo {volume} {142}},\ \bibinfo
  {pages} {1541} (\bibinfo {year} {1965})}\BibitemShut {NoStop}%
\bibitem [{\citenamefont {{Friedman}}\ and\ \citenamefont
  {{Stergioulas}}(2013)}]{FriedmanStergioulas}%
  \BibitemOpen
  \bibfield  {author} {\bibinfo {author} {\bibfnamefont {J.~L.}\ \bibnamefont
  {{Friedman}}}\ and\ \bibinfo {author} {\bibfnamefont {N.}~\bibnamefont
  {{Stergioulas}}},\ }\href@noop {} {\emph {\bibinfo {title} {{Rotating
  Relativistic Stars}}}}\ (\bibinfo  {publisher} {Cambridge University Press,
  Cambridge, UK},\ \bibinfo {year} {2013})\BibitemShut {NoStop}%
\bibitem [{\citenamefont {{Campolattaro}}\ and\ \citenamefont
  {{Thorne}}(1970)}]{Campolattaro}%
  \BibitemOpen
  \bibfield  {author} {\bibinfo {author} {\bibfnamefont {A.}~\bibnamefont
  {{Campolattaro}}}\ and\ \bibinfo {author} {\bibfnamefont {K.~S.}\
  \bibnamefont {{Thorne}}},\ }\href {\doibase 10.1086/150362} {\bibfield
  {journal} {\bibinfo  {journal} {\apj}\ }\textbf {\bibinfo {volume} {159}},\
  \bibinfo {pages} {847} (\bibinfo {year} {1970})}\BibitemShut {NoStop}%
\bibitem [{\citenamefont {{Lockitch}}\ \emph {et~al.}(2000)\citenamefont
  {{Lockitch}}, \citenamefont {{Andersson}},\ and\ \citenamefont
  {{Friedman}}}]{Lockitch_RotModesAnal}%
  \BibitemOpen
  \bibfield  {author} {\bibinfo {author} {\bibfnamefont {K.~H.}\ \bibnamefont
  {{Lockitch}}}, \bibinfo {author} {\bibfnamefont {N.}~\bibnamefont
  {{Andersson}}}, \ and\ \bibinfo {author} {\bibfnamefont {J.~L.}\ \bibnamefont
  {{Friedman}}},\ }\href {\doibase 10.1103/PhysRevD.63.024019} {\bibfield
  {journal} {\bibinfo  {journal} {\prd}\ }\textbf {\bibinfo {volume} {63}},\
  \bibinfo {pages} {024019} (\bibinfo {year} {2000})},\ \Eprint
  {http://arxiv.org/abs/gr-qc/0008019} {gr-qc/0008019} \BibitemShut {NoStop}%
\bibitem [{\citenamefont {{Thorne}}\ and\ \citenamefont
  {{Campolattaro}}(1967)}]{Thorne_Pulsations}%
  \BibitemOpen
  \bibfield  {author} {\bibinfo {author} {\bibfnamefont {K.~S.}\ \bibnamefont
  {{Thorne}}}\ and\ \bibinfo {author} {\bibfnamefont {A.}~\bibnamefont
  {{Campolattaro}}},\ }\href {\doibase 10.1086/149288} {\bibfield  {journal}
  {\bibinfo  {journal} {\apj}\ }\textbf {\bibinfo {volume} {149}},\ \bibinfo
  {pages} {591} (\bibinfo {year} {1967})}\BibitemShut {NoStop}%
\end{thebibliography}%

\end{document}